\documentclass[reprint,
 superscriptaddress,notitlepage,nofootinbib,twocolumns,
 amsmath,amssymb,
 aps,
 prx,
 showkeys
]{revtex4-1}
\usepackage[sort&compress]{natbib}
\usepackage[utf8]{inputenc}
\usepackage{graphicx}
\usepackage{braket}
\usepackage{setspace}
\usepackage{bm}
\usepackage{amsmath}
\usepackage{csquotes}
\usepackage{mathrsfs}
\usepackage{titlesec}
\usepackage{subfigure}
\usepackage[svgnames]{xcolor}
\usepackage[colorlinks,citecolor=red,urlcolor=red,bookmarks=false,hypertexnames=true]{hyperref}
\usepackage{soul}

\begin{document}

\title{Experimental observation of phase transitions of a deformed Dicke model using a reconfigurable, biparametric electronic platform}

\author{Mario A. Quiroz-Ju\'{a}rez}%
\thanks{These authors contributed equally to this work.}
\affiliation{Centro de F\'isica Aplicada y Tecnolog\'ia Avanzada, Universidad Nacional Aut\'onoma de M\'exico, Boulevard Juriquilla 3001, Juriquilla, 76230 Quer\'etaro, M\'exico}
\email{Corresponding Author: maqj@fata.unam.mx}

\author{\'{A}ngel L. Corps}
\thanks{These authors contributed equally to this work.}
\affiliation{Instituto de Estructura de la Materia, IEM-CSIC, Serrano 123, E-28006 Madrid, Spain}
\affiliation{Grupo Interdisciplinar de Sistemas Complejos (GISC), Universidad Complutense de Madrid, Avenida Complutense s/n, E-28040 Madrid, Spain}

\author{Rafael A. Molina}
\affiliation{Instituto de Estructura de la Materia, IEM-CSIC, Serrano 123, E-28006 Madrid, Spain}

\author{Armando Rela\~{n}o}
\affiliation{Grupo Interdisciplinar de Sistemas Complejos (GISC), Universidad Complutense de Madrid, Avenida Complutense s/n, E-28040 Madrid, Spain}
\affiliation{Departamento de Estructura de la Materia, Física Térmica y Electrónica, Universidad Complutense de Madrid, Avenida Complutense s/n, E-28040 Madrid, Spain}

\author{Jos\'{e} L. Aragón}%
\affiliation{Centro de F\'isica Aplicada y Tecnolog\'ia Avanzada, Universidad Nacional Aut\'onoma de M\'exico, Boulevard Juriquilla 3001, Juriquilla, 76230 Quer\'etaro, M\'exico}

\author{Roberto de J. Le\'on-Montiel}
\affiliation{Instituto de Ciencias Nucleares, Universidad Nacional Aut\'onoma de M\'exico, Apartado Postal 70-543, 04510 Cd. Mx., M\'exico}

\author{Jorge G. Hirsch}
\affiliation{Instituto de Ciencias Nucleares, Universidad Nacional Aut\'onoma de M\'exico, Apartado Postal 70-543, 04510 Cd. Mx., M\'exico}

\date{\today}

\begin{abstract}
We experimentally study the infinite-size limit of the Dicke model of quantum optics with a parity-breaking deformation strength that couples the system to an external bosonic reservoir. We focus on the dynamical consequences of such symmetry-breaking, which makes the classical phase space asymmetric with non-equivalent energy wells. We present an experimental implementation of the classical version of the deformed Dicke model using a state-of-the-art bi-parametric electronic platform. Our platform constitutes a playground for studying representative phenomena of the deformed Dicke model in electrical circuits with the possibility of externally controlling parameters and initial conditions. In particular, we investigate the dynamics of the ground state, various phase transitions, and the asymmetry of the energy wells as a function of the coupling strength $\gamma$ and the deformation strength $\alpha$ in the resonant case. Additionally,  to characterize the various behavior regimes, we present a two-dimensional phase diagram as a function of the two intrinsic system parameters. The onset of chaos is also analyzed experimentally. Our findings provide a clear connection between theoretical predictions and experimental observations, demonstrating the usefulness of our bi-parametric electronic setup.

\end{abstract}

\maketitle

\section{INTRODUCTION} 

Our understanding of the physical world is largely reliant on the concept of phase. Physical systems exhibit phases of qualitatively different nature. These may be characterized by some thermodynamic and dynamical properties which in turn allow us to define what those phases are. In recent years, phase transitions have played a paramount role in the discovery of exotic effects and, in particular, in the field of quantum technologies \cite{Baumann2010,Muniz2020}. The simplest example of a phase transition is the traditional ground-state quantum phase transition (QPT), whereby the lowest energy of a system undergoes an abrupt change when some control parameters of the model are varied \cite{sachdev2011}. An extension of the QPT concept to the high-lying states of a system has been the subject of much recent research: the excited-state quantum phase transition (ESQPT) \cite{cejnar2021}. ESQPTs entail certain non-analyticities in the spectrum of a system such as some divergences in its density of states at certain critical energies. The phases of a system may also be described by dynamical effects such as the development of chaos, which implies the loss of information about the exact initial state and the practical unpredictability of what the future evolution of a given state will be. These notions (with some restrictions) can be realized both in the quantum and classical realms. Such quantum-classical correspondence may be readily established for quantum systems with a semiclassical counterpart, which is usually achieved through some sort of infinite-size limit of the quantum model. It turns out that the Dicke \cite{Dicke1954} and related quantum optical models, originally devised as toy-models to describe the light-matter interaction, are exceptionally useful to analyze non-trivial phase-transition phenomena \cite{corps2021PRA,Corps2021PRL,Bastarrachea2014a,Puebla2013,Relano2008,Stransky2021,Hwang2015,Puebla2016,Klinder2015} and related concepts such as chaos and thermalization \cite{MurPetit2018,Relano2011,Bastarrachea2014b,LewisSwan2019,Cameo2021,Cameo2020,Relano2018,Kloc2018,ChavezCarlos2019,Lobez2016,corps2022chaos,Emary2003,wang2022}. Thus, they have become a powerful testbed for the study of quantum statistical mechanics \cite{larson2021book}.

The Dicke model represents the interaction of an ensemble of atoms (matter) with a monochromatic harmonic field (light), the intensity of such interaction being mediated by a parameter, $\gamma$.  It is a collective (or fully-connected) system as all of its constituents interact simultaneously. The collective nature of the family of atom-field models that the Dicke model belongs to allows that many \cite{larson2017} relevant physical quantities of the quantum model can be understood from its infinite-size limit \cite{cejnar2021,Corpscomment}. These quantities notably include phase transitions as well as ground-state properties. As the system size increases, the model approaches classicality. A deformed version of the Dicke model was introduced in \cite{corps2022chaos}, where the consequences of an additional symmetry-breaking term, $\alpha$, were analyzed. Such symmetry-breaking is achieved through a deformation strength that couples the traditional light-matter interaction to an external bosonic reservoir. From the viewpoint of its classical counterpart, the intensity of the deformation is responsible for the appearance of asymmetric, non-equivalent energy wells. Within each well, much of the dynamics takes place irrespective of the dynamical properties of the other well. For example, the transition to chaos occurs at different energies for each of the wells. This is so because in the infinite-size limit such classical wells are separated by an infinite potential barrier and no tunneling is possible. 


In this work, we present an experimental implementation of the classical version of the deformed Dicke model. The main purpose is to explore experimentally the effect that the asymmetric wells have on the system dynamics as well as the transition from two equivalent wells to a single deformed well. We are interested in the role played by electronic noise, which serves to account for the imperfections in a real physical system. In our experimental implementation, we make use of a state-of-the-art bi-parametric electronic platform that comprises passive and active elements. Our electronic platform allows us to investigate the dynamics of the ground state, various phase transitions, and the asymmetry of the energy wells, all as a function of two intrinsic system parameters ($\gamma,\alpha$) in the resonant case. We utilize numerical techniques to establish a clear connection between the theoretical predictions and experimental observations.

The structure of the paper is as follows. In Sec. \ref{sec:model}, we briefly review the parity-broken Dicke model and its representation in electrical variables of LC oscillators. In Sec.  \ref{sec:platform} we describe the electronic implementation of the semiclassical approximation of the deformed Dicke model. In Sec. \ref{sec:results}, we describe experimental measurements of some representative phenomena of the model, which were carried out in our bi-parametric electronic platform. We finally present our conclusions in Sec. \ref{sec:conclusion}.

\section{MODEL}
\label{sec:model}

We consider the modified version of the standard Dicke model introduced in \cite{corps2022chaos}, which describes the interaction between $N$ identical two-level atoms and a single-mode quantized radiation field, including a direct coupling to an external bosonic reservoir represented by a deformation strength. This model is represented by the Hamiltonian

\begin{equation}
    \mathcal{H}=\omega \hat{a}^{\dagger}\hat{a}+\omega_0 \hat{J}_z+\left(\frac{2\gamma}{\sqrt{N}}\hat{J}_x+\sqrt{\frac{N\omega_{0}}{2}}\alpha\right)(\hat{a}^{\dagger}+\hat{a}).
    \label{eq:Hquantum}
\end{equation}
where  $\hat{a}^{\dagger}$ and $\hat{a}$ are the creation and annihilation Bose operators of the quantized radiation field with frequency $\omega$. For convenience, we have set $\hbar=1$. The collective pseudo-spin operators, $\hat{\mathbf{J}}=(\hat{J_x}, \hat{J_y},\hat{J_z})$, $\hat{J}_{k}=\frac{1}{2}\sum_{i=1}^{N}\hat{\sigma}_{i}^{k}$ ($k=x,y,z$), with $\hat{\sigma}_{i}^{k}$ being the Pauli matrices acting on atom $i$, obey the $SU(2)$ algebra and represent the $N$ two-level atoms, each with a separation energy of $\omega_0$. The total angular momentum operator $\hat{\mathbf{J}}^2=\hat{J}_{x}^{2}+\hat{J}_{y}^{2}+\hat{J}_{x}^{2}$ is a conserved quantity with eigenvalues $j(j+1)$, whose symmetric atomic subspace, including the ground state, is defined by the maximum value $j=N/2$. The parameter $\gamma$ represents the atom-field coupling strength, which depends on the atomic dipole moment, and $\alpha$ represents the deformation strength. The case $\alpha=0$ yields the traditional Dicke model. The parity symmetry of the standard Dicke model, $\Pi=e^{i\pi(j+\hat{J}_{z}+\hat{a}^{\dagger}\hat{a})}$, is broken when $\alpha \neq0$, i.e., $[{\mathcal{H}},{\Pi}]\neq 0$, leading to quantitative changes \cite{corps2022chaos}. When $\alpha=0$, the ground state of the system undergoes a normal-superradiant phase transition at the critical value of the coupling strength $\gamma_c=\sqrt{\omega \omega_0}/2$. Another notable phenomenon for $\alpha=0$ is the existence of an ESQPT for $\gamma>\gamma_c$ at the 
energy $E=-j \omega_0$ \cite{Bastarrachea2014a,PerezFernandez2011,Brandes2013}. In what follows, we will let $E$ denote the energy normalized by $j\omega_{0}$ for convenience. For $\alpha\neq0$, there is no normal-superradiant phase transition as the population of the ground-state is positive for all values of $\gamma$, and various ESQPTs appear at different values of the excitation energy whose value, however, cannot be expressed in terms of elementary functions \cite{corps2022chaos}.

A semiclassical approximation of the deformed Dicke Hamiltonian (\ref{eq:Hquantum}), in the thermodynamic limit $N \rightarrow \infty$, can be calculated by taking the expectation value of the Hamiltonian operator in the tensor product of Glauber and Bloch coherent states \cite{de1992chaos, chavez2016classical}. These states are defined as $\left|\beta\right>=e^{-\left|\beta\right|^2/2}e^{\beta \hat{a}^{\dagger}}\left|0\right>$ and $\left|z\right>=1/(1+\left|z\right|^2)^{-j}e^{z\hat{J}_+}\left|j,-j\right>$, respectively, where $\left|0\right>$ is the photon vacuum and $\left|j,-j\right>$ is the ground state for the atoms. This semiclassical approximation provides a mean-field solution that captures important characteristics of the quantum model in the infinite-size limit. By employing the real canonical variables $(q,p)$ for the field and $(Q,P)$ for the atomic part, expressed in terms of the coherent states with $\beta=\sqrt{j/2}(q+ip)$ and $z=\frac{(Q-iP)}{\sqrt{4-Q^2-P^2}}$, the classical analogue per particle of the modified Dicke Hamiltonian (\ref{eq:Hquantum}), $\left<\beta,z\right| \mathcal{H} \left|\beta,z\right>/\omega_0 j$ reads,
\begin{eqnarray}\label{eq:hamclassical}
    \small H= &\frac{\omega}{2\omega_0}(q^2+p^2)+\frac{1}{2}(Q^2+P^2)+ \nonumber \\
    &\frac{\gamma q Q}{\omega_0}\sqrt{4-P^2+Q^2}-1+\sqrt{\frac{2}{\omega_0}}\alpha q.
    \label{eq:Hclassical}
\end{eqnarray}
The classical phase space is $\mathcal{M}=\mathbb{R}^{2}\times\mathbb{S}$ as the photonic variables are unbouded, $q,p\in\mathbb{R}$,  but the atomic ones are restricted to a two-dimensional ball of radius 2, $Q^{2}+P^{2}\leq 4$. In the infinite-size limit, the effective Planck constant vanishes, $\hbar_{\textrm{eff}}\propto 1/N\to 0$, leading to classicality. To explore the dynamical properties of the model (\ref{eq:Hclassical}), we resort to the Hamilton equations of motion:

\begin{eqnarray}
\dot{q}&=&\frac{\omega}{\omega_0} p,\nonumber \\
\dot{p}&=&-\frac{\gamma Q}{\omega_0}\sqrt{4-P^2-Q^2}-\frac{\omega}{\omega_0}q-\sqrt{\frac{2}{\omega_0}}\alpha, \nonumber\\
\dot{Q}&=&P-\frac{\gamma PqQ}{\omega_0\sqrt{4-P^2-Q^2}},\nonumber \\
\dot{P}&=&\frac{\gamma q Q^2}{\omega_0\sqrt{4-P^2-Q^2}}-\frac{\gamma q}{\omega_0} \sqrt{4-P^2-Q^2}-Q.
\label{eq:motionpq}
\end{eqnarray}

\begin{figure*}[t]
\includegraphics[width=18cm]{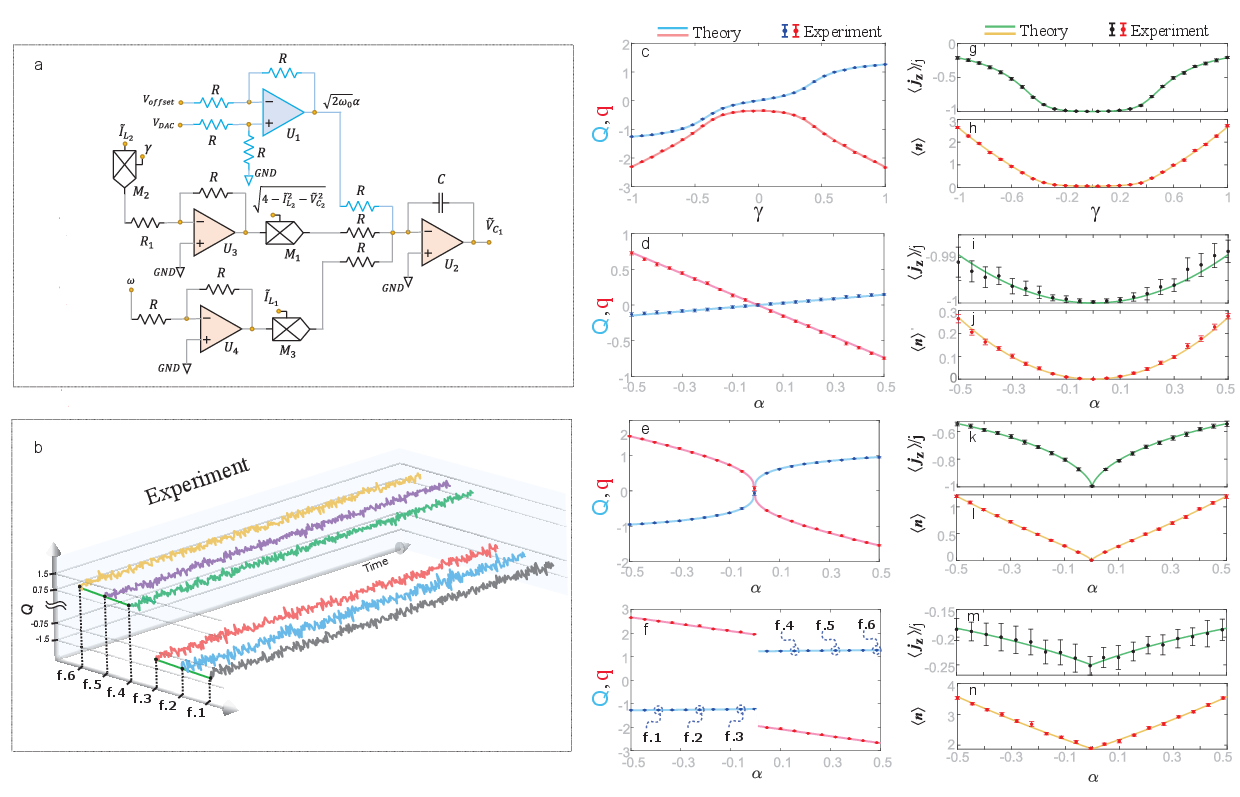}
\caption{(\textbf{a}) Scheme of the modified electronic circuit that describes the differential equation of the voltage in the capacitor-1, or equivalently, $\dot{p}$ in Eq. (\ref{eq:motionpq}). The circuit in blue color indicates the electronic components added to the circuit schematic diagram proposed in \cite{quiroz2020experimental}. Here, $U_j$ and $M_j$ stand for general-purpose
operational amplifiers and analog multipliers. The resistor and capacitor values used in the circuit
are the following: $R=$10 $K\Omega$, $R_1=$1 $K\Omega$, and $C=$10 $\mu F$. V$_{\textrm{offset}}$ is a direct voltage of 2.5 volts and V$_{\textrm{DAC}}$ is the voltage signal provided by the DAC, which depends on the value of $\alpha$. (\textbf{b}) Temporal evolution of $Q$ coordinate in the ground-state for $\gamma=1$ and different values of $\alpha$: (\textbf{f.1}) $\alpha=-0.39$, (\textbf{f.2}) $\alpha=-0.225$, (\textbf{f.3}) $\alpha=-0.06$, (\textbf{f.4}) $\alpha=0.155$, (\textbf{f.5}) $\alpha=0.32$, and (\textbf{f.6}) $\alpha=0.485$. (\textbf{c})-(\textbf{f}) Coordinates $q$ and $Q$ of the ground state corresponding to the classical version of the deformed Dicke model on resonance, $\omega=\omega_0=1$, as a function of the coupling strength $\gamma$ and deformation strength $\alpha$. The free parameters in each case are set to: (\textbf{c}) $\alpha=0.25$, (\textbf{d}) $\gamma=0.1<\gamma_{c}$, (\textbf{e}) $\gamma=\gamma_{c}=0.5$ and (\textbf{f}) $\gamma=1>\gamma_{c}$. Figures (\textbf{g})-(\textbf{i})-(\textbf{k})-(\textbf{m}) show the classical analogous of the atomic inversion,  $\left<j_z\right>/j$, and (\textbf{h})-(\textbf{j})-(\textbf{l})-(\textbf{n}) the classical analogous of the mean photon number $\left<n\right>$ as a function of $\gamma$ and $\alpha$. (\textbf{g})-(\textbf{h}) correspond to the case $\alpha=0.25$, (\textbf{i})-(\textbf{j}) to $\gamma=0.1$, (\textbf{k})-(\textbf{l}) to $\gamma=0.5$, and (\textbf{m})-(\textbf{n}) to $\gamma=1$. Solid lines denote theoretical predictions obtained via numerical simulations and points represent experimental measurements. Error bars correspond to the standard deviation of the experimental results.}
\label{fig:figure1}
\end{figure*}

The classical Hamiltonian (\ref{eq:Hclassical}) describes two non-linearly coupled harmonic oscillators with an external stimulus in the field section. It can be rewritten in terms of the electrical variables of LC oscillators, where L and C stand for inductance and capacitance, respectively. This enables an electronic version of the system to be implemented, as previously demonstrated for the standard Dicke model \cite{quiroz2020experimental}. The transformation results in a system of two non-autonomous LC oscillators with a nonlinear coupling. Now, we focus on mapping the canonical variables $(q,p,Q,P)$ to the normalized electrical variables $({I}_{L_1}, {V}_{C_1}, {I}_{L_2}, {V}_{C_2})$. Here, ${I}_{L}$ represents the current in the inductor and ${V}_{C}$ represents the voltage in the capacitor. For more details, see Ref. \cite{quiroz2020experimental}. In terms of these new variables, the classical Hamiltonian (\ref{eq:Hclassical}) becomes

\begin{eqnarray}
H&=&\frac{\omega}{2\omega_0}(\tilde{I}_{L_1}^2+\tilde{V}_{C_1}^2)+\frac{1}{2}(\tilde{I}_{L_2}^2+\tilde{V}_{C_2}^2) \nonumber \\&+&\frac{\gamma \tilde{I}_{L_1} \tilde{I}_{L_2}}{\omega_0}\sqrt{4-\tilde{V}_{C_2}^2+\tilde{I}_{L_2}^2}-1+\sqrt{\frac{2}{\omega_0}}\alpha \tilde{I}_{L_1},
\label{eq:HclassicalLC}
\end{eqnarray}
where $\omega^2 = 1/L_1C_1$ and $\omega_0^2= 1/L_2C_2$ are the natural frequencies of the oscillator-1 and oscillator-2, respectively.  Note that the coupling strength $\gamma$ controls the nonlinear interaction between the LC oscillators and $\alpha$ governs the intensity of the external stimulus. In 
the experimental setup we set $\omega=1$ and $\omega_0=1$ and study the dependence of the observables on the $\gamma$ and $\alpha$ coefficients. 

\section{BI-PARAMETRIC ELECTRONIC PLATFORM}
\label{sec:platform}

The standard Dicke model has been experimentally explored in its quantum version using cold atoms \cite{dimer2007proposed, kongkhambut2021realization}, and superconducting circuits \cite{gonzalez2013mesoscopic, mlynek2014observation}. Moreover, several extensions of the model have been proposed and implemented using atom-cavity systems \cite{zhiqiang2017nonequilibrium, masson2017cavity}. However, experimental realizations of the classical version of the Dicke model are scarce, despite its extensive study. Thus, the design of physical platforms to implement this model remains an important area of research as it can lead to the verification of key predictions of the model and provide new insights into its properties. 
Reference \cite{quiroz2020experimental} presented the first experimental implementation of the classical Dicke model. This was managed by means of two non-linearly coupled synthetic LC circuits. The experimental setup was based on electrical networks of resistors, capacitors, operational amplifiers (OPAMPs), and analog multipliers \cite{carlson1967handbook, johnson1963analog}. The remarkable feature of these electrical networks is that their voltage transfer-functions correspond to linear and nonlinear operations, thus enabling the definition of differential equations. As a result, the temporal evolution of the voltages in the electronic device is governed by the same equations as the original physical system \cite{noordergraaf1963use, moore1998dynamical}. This approach has been found to be effective in simulating dynamical systems in real time \cite{vazquez2013arbitrary, quiroz2019generation, jimenez2021experimental, escobar2022classical, chen2020effect} and investigating phenomena requiring high control of parameters \cite{leon2015noise, quiroz2017emergence, leon2018observation, quiroz2021reconfigurable, quiroz2021demand}. The use of electric circuits has also been of notable interest in other kinds of problems, such as in topological systems \cite{Dong2021,Hua2018,imhof2018topolectrical,albert2015topological}. It is worth mentioning the first implementation of Chua's circuit, the simplest electronic circuit exhibiting chaos, experimentally confirmed in \cite{Zhong1985} (see also \cite{Kuznetsov2023}).

\begin{figure}[t]
\includegraphics[width=8.5cm]{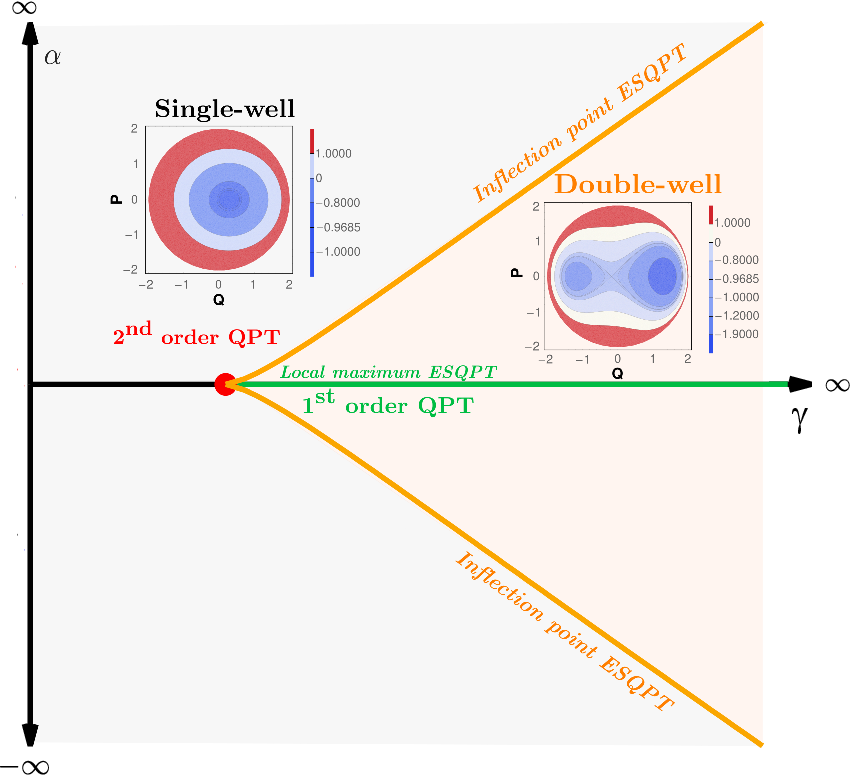}
\caption{Schematic phase diagram of the deformed Dicke model as a function of $\gamma$ and $\alpha$. A second-order (ground-state) QPT takes place at $\alpha=0$, $\gamma_{c}=\sqrt{\omega \omega_{0}}/2$. For each $\gamma>\gamma_{c}$, there exists a critical value of $\alpha$, $\alpha_{c}$, such that if the deformation becomes too strong, $\alpha>\alpha_{c}$, the classical system only exhibits a single energy well, while if $\alpha<\alpha_{c}$ the system may exhibit up to two classical wells, which are connected by an ESQPT at a given critical energy. The values $\alpha_{c}(\gamma)$ are shown with orange curves obtained numerically from the semiclassical model. Crossing the $\alpha=0$ axis when $\gamma>\gamma_{c}$ gives rise to a first-order QPT. Representative orbits in phase space are presented both for the single- and double-well scenarios, obtained as the projection on the $(Q,P)$ plane of the four-dimensional contour surfaces of the classical Hamiltonian \eqref{eq:hamclassical} at a given energy (see colorbars).}
\label{fig:figure2}
\end{figure}

\begin{figure*}[t]
\includegraphics[width=16.5cm]{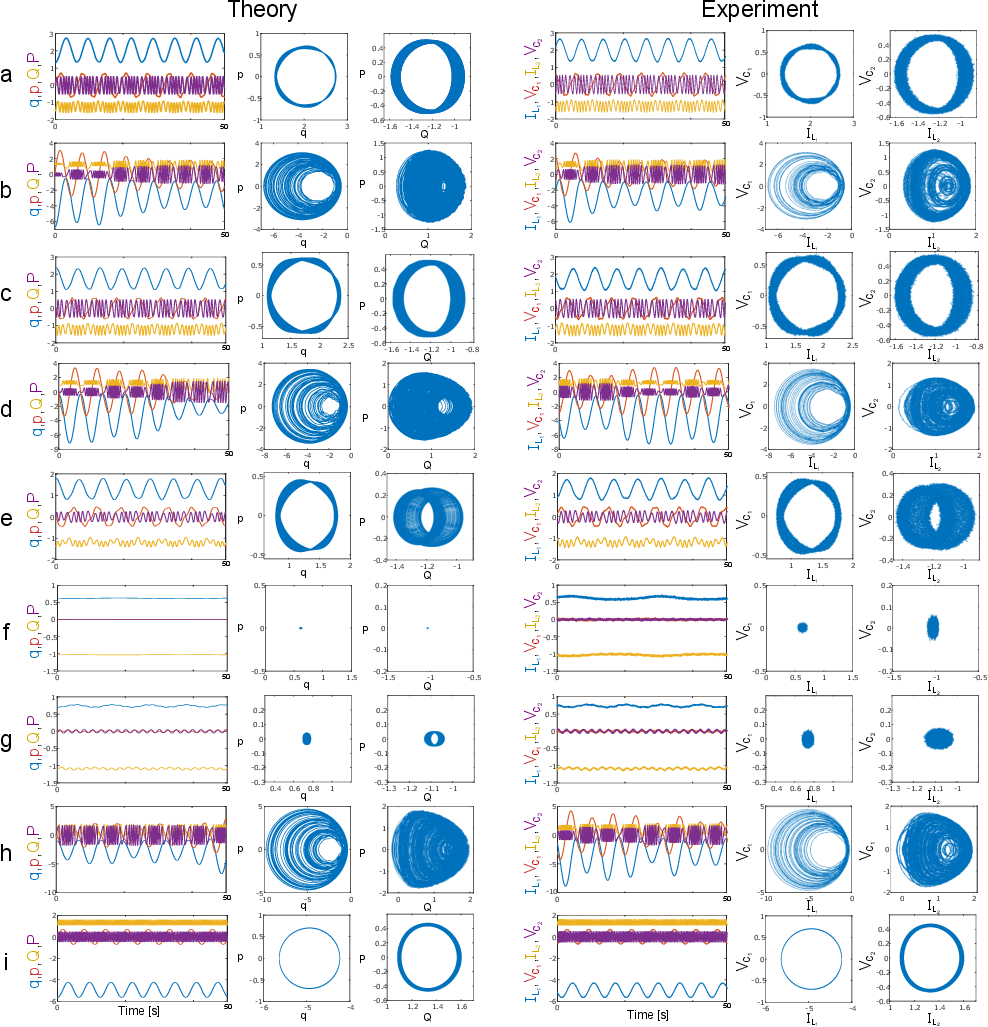}
\caption{Left-hand side shows the temporal evolution of the coordinates ($q,p,Q,P$) and the projections of the trajectories in the planes ($q,p$) and ($Q,P$) for numerical simulations, and the right-hand side presents the time evolution of the electrical coordinates ($I_{L_1},V_{C_1},I_{L_2},V_{C_2}$) and their corresponding projections in the planes ($I_{L_1},V_{C_1}$) and ($I_{L_2},V_{C_2}$) for the experimental results, considering different energy and deformation strength values: (\textbf{a}) $E=-1.9993$ and $\alpha=0.5$, (\textbf{b}) $E=-2.0007$ and $\alpha=0.5$, (\textbf{c}) $E=-1.4997$ and $\alpha=0.7$, (\textbf{d}) $E=-1.4984$ and $\alpha=0.7$, 
(\textbf{e}) $E=-0.9998$ and $\alpha=1.1$, (\textbf{f}) $E=-0.6830$ and $\alpha=1.41$, (\textbf{g}) $E=-0.6657$ and $\alpha=1.43$, (\textbf{h}) $E=-2.0001$ and $\alpha=1.5$ and, (\textbf{i}) $E=-12.0001$ and $\alpha=1.5$. In each case, the initial conditions ($q,p,Q,P$) are set to: (\textbf{a}) $(1.32,0,-1.5,0)$, (\textbf{b}) $(-6.736,0,1.5,0)$, (\textbf{c}) $(2.374,0,-1,0)$, (\textbf{d}) $(-0.29,0,1,0)$, (\textbf{e}) $(1.797,0,-1.4,0)$, (\textbf{f}) $(0.694,0,-1.1,0)$, (\textbf{g}) $(0.608,0,-1.019,0)$, (\textbf{h}) $(-1,0,-0.201,0)$ and (\textbf{i}) $(-5.638,0,1.1,0)$. We set $\gamma=1.5$, $\omega=1$ s$^{-1}$ and $\omega_0=1$ s$^{-1}$ for all cases.}
\label{fig:figure3}
\end{figure*}

An important feature of the electronic platform proposed in \cite{quiroz2020experimental} is its simplicity and versatility in controlling the system's parameters and initial conditions via external voltage signals. In this work, we exploit this feature to experimentally investigate the phase transitions in the semiclassical approximation of the deformed Dicke model using a modified version of the electronic platform. 

A crucial observation is that the classical motion equations of the standard Dicke model and the equations from the deformed Dicke model differ only in the exogenous term $-\sqrt{2/\omega_0}\alpha$ in $\dot{p}$. 
It can be introduced by including a direct voltage signal in the integrator for the differential equation that describes the voltage of the capacitor in oscillator-1 ($\tilde{V}_{C_1}$), or equivalently in the differential equation for $\dot{p}$, to represent the exogenous term. This input voltage is provided by a digital-analog converter (DAC) with an MCP4921 series, and its magnitude depends on the value of $\alpha$. A subtractor amplifier is also included before the DAC to fix an offset level of 2.5 volts, allowing for values of $\alpha$ less than zero, i.e., the product $\sqrt{2}\alpha$ can take values in the interval (-2.5, 2.5). A similar strategy is used to extend the range of variation for the coupling strength $\gamma\in(-2.5,2.5)$. In this way the system has two controlable free parameters, $\gamma$ and $\alpha$, which are set via software using a serial peripheral interface protocol to communicate with the DACs and a master 8-bit microcontroller. The electronic circuit for the deformed Dicke model is shown in Fig. \ref{fig:figure1}(\textbf{a}). Note that the black-colored schematics represent the circuitry for the standard Dicke model, while the blue schematics represents the exogenous term that deforms it. More details about the experimental setup are provided in the Appendix.

Our setup provides a biparametric reconfigurable electronic platform synthesized with analog electrical components that enables the experimental exploration of phase transitions and characteristic phenomena exhibited by the classical version of the deformed Dicke model. We use a Tektronix TBS200 oscilloscope to collect 300 second time series of the voltage in the capacitor and the current through the inductor in each LC oscillator, to track the temporal evolution of the classical Hamiltonian (\ref{eq:HclassicalLC}) under different conditions.

\section{EXPERIMENTS}
\label{sec:results}

In the following, we present the results of a series of experiments conducted on our bi-parametric electronic platform, which implements the semiclassical approximation of the deformed Dicke model. Our focus is on the study of various characteristic phenomena, including the determination of the ground-state and its phase transitions, as well as the dynamics of parity-symmetry breaking.

\subsection{Ground state}

The ground state can be obtained by evaluating the Hamiltonian (\ref{eq:Hclassical}) at the stable equilibrium points, and selecting the fixed point with minimum energy. When $\alpha=0$, there are analytical closed-form expressions for the fixed points, and the ground-state energy is given by $E_0(\gamma)=-\omega_0$ 
for $\gamma\leq\gamma_c$ and $E_0(\gamma)=-\omega_0/2(\gamma_c^2/\gamma^2+\gamma^2/\gamma_c^2)$ for $\gamma>\gamma_c$. The critical point $\gamma_c=\sqrt{\omega \omega_0}/2$ signals the transition from normal to superradiant phase.  It is worth noting, as described in Ref. \cite{corps2022chaos}, that the Hamiltonian (\ref{eq:Hclassical}) for $\alpha=0$ is symmetric under $q \rightarrow -q$ and $Q \rightarrow -Q$, so $\mathbb{X}_{+}=(q,p,Q,P)$ and $\mathbb{X}_{-}=(-q,p,-Q,P)$ are fixed points with the same energy.  These points satisfy the dynamical equations (\ref{eq:motionpq}) when the temporal derivatives are zero, leading to a ground-state that is degenerate and has two symmetric global energy minima. However, when $\alpha\neq0$, the scenario changes dramatically. Firstly, it is not possible to find a closed expression for the critical points, in consequence, the analysis is restricted to a numerical treatment. Second, the ground state no longer is degenerate and two asymmetric wells may emerge at different energies (see below).

We have carried out numerical simulations to obtain the ground state of the classical deformed Dicke model by fixing one of the two free parameters and varying the other. Once the fixed points with energy minima have been identified for different values of the parameter, we track them experimentally to validate our bi-parametric electronic platform.  Figures \ref{fig:figure1}(\textbf{c})-(\textbf{f}) show the coordinates $q$ and $Q$ of the ground state, on resonance $\omega=\omega_0=1$, considering in each case the following  fixed parameters: (\textbf{c}) $\alpha=0.25$, (\textbf{d}) $\gamma=0.1<\gamma_{c}$,  (\textbf{e}) $\gamma=0.5=\gamma_{c}$, and (\textbf{f}) $\gamma=1>\gamma_{c}$. The blue and red solid lines are the corresponding theoretical predictions for $q$ and $Q$ coordinates, respectively, whereas the blue and red points are experimental measurements obtained from our electronic platform. In the classical case, the ground state is represented by a stable critical point with a stationary solution. In our bi-parametric electronic platform, this solution is traduced as a constant voltage signal along the time. Electronic noise inherently affects the constant voltage signal, preventing it from remaining perfectly flat. Figure \ref{fig:figure1}(\textbf{f}) shows the temporal evolution of the $Q$ coordinate (with fixed parameter $\gamma=1$), where stochastic fluctuations can be observed due to the inherent noise present in the electronic platform for different values of $\alpha$, (\textbf{f.1}) $\alpha=-0.39$, (\textbf{f.2}) $\alpha=-0.225$, (\textbf{f.3}) $\alpha=-0.06$, (\textbf{f.4}) $\alpha=0.155$, (\textbf{f.5}) $\alpha=0.32$, and (\textbf{f.6}) $\alpha=0.485$. The error bars in Figs. \ref{fig:figure1}\textbf{(c)}-\textbf{(f)} show the standard deviations of the noisy time series for each of the $(q,Q)$ coordinates, with the blue and red points representing their respective mean values. Moreover, for the cases studied in Figs. \ref{fig:figure1}(\textbf{c})-(\textbf{f}), we calculate the classical analogue of atomic inversion \begin{equation} \frac{\left<j_z\right>}{j}=\frac{\bra{\beta,z}\hat{J}_{z}\ket{\beta,z}}{j}=\frac{P^{2}+Q^{2}}{2}-1,
\end{equation}
and the mean photon number,
\begin{equation}\left<n\right>=\frac{p^{2}+q^{2}}{2},
\end{equation}
as a function of $\gamma$ and $\alpha$, as shown in Figs. (\textbf{g}), (\textbf{i}), (\textbf{k}), (\textbf{m}) and
Figs. (\textbf{h}), (\textbf{j}), (\textbf{l}), (\textbf{n}), respectively. Figures (\textbf{g})-(\textbf{h}) correspond to the case $\alpha=0.25$, (\textbf{i})-(\textbf{j}) to $\gamma=0.1$, (\textbf{k})-(\textbf{l}) to $\gamma=0.5$, and (\textbf{m})-(\textbf{n}) to $\gamma=1$. In all cases, the solid lines describe the  theoretical predictions and the points represent the experimental results.

When $\alpha$ is fixed, the $q$ and $Q$ coordinates as a function of the coupling strength $\gamma$ vary smoothly as $\gamma$ increases [see Fig. \ref{fig:figure1}(\textbf{c})]. This transition is verified by the atomic inversion in Fig. \ref{fig:figure1}(\textbf{g}) and the mean photon number in Fig. \ref{fig:figure1}(\textbf{h}). This is because as $\gamma$ increases from $\gamma=0$, when there are no interactions and the ground state is found at the center of the phase space ($Q=0$), the system will eventually enter the double-well region at a certain value of $\alpha$, and thus the ground-state is shifted to the right ($Q>0$). Analogous results are found when $\gamma$ is decreased from $\gamma=0$. When the coupling strength is set to $\gamma=0.1<\gamma_{c}$ and it is $\alpha$ that varies, the ground state exhibits a monotonic behavior, as shown in Fig. \ref{fig:figure1}(\textbf{d}) for both $q$ and $Q$ coordinates, and Figs. \ref{fig:figure1}(\textbf{i}) and (\textbf{j}) for the atomic inversion and mean photon number, respectively. The reason is that for $\gamma<\gamma_{c}$, there is always a single classical well, but it is shifted to the left ($Q<0$) or right ($Q>0$) depending on the sign of $\alpha$, only becoming centered ($Q=0$) when there is no deformation, $\alpha=0$. Figure \ref{fig:figure1}(\textbf{e}) displays the ground state values as a function of $\alpha$. At $\gamma=0.5=\gamma_{c}$ and $\alpha=0$, the system experiences a sudden change in the values of $q$ and $Q$ coordinates, indicating a second-order quantum phase transition. This phenomenon is also reflected in the atomic inversion shown in Fig. \ref{fig:figure1}(\textbf{k}) and the mean photon number presented in Fig. \ref{fig:figure1}(\textbf{l}). These figures show that, as the deformation strength approaches the critical value $\alpha \rightarrow 0$, the atomic inversion approaches $-1$ and the mean photon number vanishes. Indeed, the point $(\gamma_{c},\alpha=0)$ corresponds to the well-known second-order QPT of the Dicke model, where there is phase coexistence of two symmetric wells. However, for $\alpha\neq 0$, the deformation forces the classical phase space to become asymmetric, which means that the ground state is shifted either to the right or left region. Finally, for the case $\gamma=1>\gamma_{c}$, the ground state exhibits a first-order quantum phase transition, characterized by a discontinuity of the canonical $q,Q$ at $\alpha=0$ for $\gamma>\gamma_{c}$, as shown in Fig. \ref{fig:figure1}(\textbf{f}). The atomic inversion in Fig. \ref{fig:figure1}(\textbf{m}) and mean photon number in Fig. \ref{fig:figure1}(\textbf{n}) illustrate the nature of the phase transition at $\alpha=0$. In this sense, the canonical variables $q,Q$ signal the different phase transitions of the model, which are then reflected in derived quantities such as the atomic inversion and the mean photon number, although they are not order parameters properly speaking. In general, there is excellent agreement between the measurements and the expected numerical results.

\subsection{Phase transitions and chaos}

To characterize the behavior regimes of the classical deformed Dicke model depicted in the previous section, we present a two-dimensional phase diagram as a function of $\gamma$ and $\alpha$ (see Fig. \ref{fig:figure2}), which fully describes the behavior of the deformed Dicke model in the thermodynamic limit, $N\to\infty$. This diagram is divided into two zones by orange solid lines. Notably, these lines originate at the critical value $\gamma_c$ where the well-known second-order quantum phase transition of the standard Dicke model ($\alpha=0$) takes place. In the gray zone, the system exhibits behaviors confined in a single well. However, in the orange-shaded region the system presents two energy wells, so the projections of the trajectories in the plane $(q,p)$ and $(Q,P)$ are formed by two lobes. The transition from single-well to double-well occurs at a critical value of $\alpha$ for each fixed $\gamma$, $\alpha_{c}=\alpha_{c}(\gamma)$, signaled by the orange line. For $\alpha=0$, the Hamiltonian admits two symmetric global energy minima, whereas for $\alpha\neq 0$ the parity symmetry is broken and two asymmetric energy wells may appear depending on $\gamma$ \cite{corps2022chaos}. Interestingly, the sign of $\alpha$ controls the asymmetry of the wells. The location in phase space of the ground-state of the system depends on the sign of $\alpha$: if $\alpha>0$, then the ground-state has $Q>0$, while if $\alpha<0$ it has $Q<0$. The second classical well appears at a higher energy and on the opposite side of classical phase space.  Additionally, the classical deformed Dicke model exhibits an excited state quantum phase transition, which occurs at a given excited energy only for $\alpha<\alpha_{c}(\gamma)$ and $\gamma\geq \gamma_{c}$, within the orange shaded region. This ESQPT is responsible for the merging of the two previously disconnected classical wells \cite{corps2022chaos}. 

We demonstrate the versatility of our electronic platform by performing experiments with different energies and deformation strengths, (\textbf{a}) $E=-1.9993$ and $\alpha=0.5$, (\textbf{b}) $E=-2.0007$ and $\alpha=0.5$, (\textbf{c}) $E=-1.4997$ and $\alpha=0.7$, (\textbf{d}) $E=-1.4984$ and $\alpha=0.7$,
(\textbf{e}) $E=-0.9998$ and $\alpha=1.1$, (\textbf{f}) $E=-0.6830$ and $\alpha=1.41$, (\textbf{g}) $E=-0.6657$ and $\alpha=1.43$, (\textbf{h}) $E=-2.0001$ and $\alpha=1.5$ and, (\textbf{i}) $E=-12.0001$ and $\alpha=1.5$,  while keeping the coupling strength constant at $\gamma=1.5$. The left-hand side of Fig. \ref{fig:figure3} shows the temporal evolution of the coordinates ($q$, $p$, $Q$, $P$), as well as the projections of the trajectories in the planes $(q, p)$ and $(Q, P)$ for the theoretical predictions. On the right-hand side of the same figure, we present the corresponding electrical coordinates ($I_{L_1}$, $V_{C_1}$, $I_{L_2}$, $V_{C_2}$), and their respective projections in the planes ($I_{L_1}$, $V_{C_1}$) and ($I_{L_2}$, $V_{C_2}$). In each case, the initial conditions ($q,p,Q,P$) are set to: (\textbf{a}) $(1.32, 0, -1.5, 0)$, (\textbf{b}) $(-6.736,0,1.5,0)$, (\textbf{c}) $(2.374,0,-1,0)$, (\textbf{d}) $(-0.29,0,1,0)$, (\textbf{e}) $(1.797,0,-1.4,0)$, (\textbf{f}) $(0.694,0,-1.1,0)$, (\textbf{g}) $(0.608,0,-1.019,0)$, (\textbf{h}) $(-1,0,-0.201,0)$ and (\textbf{i}) $(-5.638,0,1.1,0)$. 
For $\gamma=1.5$, the critical value of the deformation strength separating the single- and double-well phases is $\alpha_c \approx 1.43019$. In this study, we analyze three deformation strengths lower than $\alpha_c$, as shown in Figs. \ref{fig:figure3}(\textbf{a})-(\textbf{e}). Note that cases (\textbf{a}) and (\textbf{b}) have the same values of energies and $\alpha$. Under this parameter configuration, the system presents two disconnected wells, located at opposite sides of the phase space. We examine two different initial conditions to explore the dynamics of each well. Fig. \ref{fig:figure3}(\textbf{a}) shows that the right well exhibits periodic trajectories in the phase plane, while Fig. \ref{fig:figure3}(\textbf{b}) demonstrates that the left well has chaotic dynamics, as can be visualized because the trajectory densely covers the available phase space. Furthermore, the available evolution space in the left well is larger than that in the right well, highlighting the asymmetry induced by the deformation strength. This is in good agreement with the theoretical analysis presented in \cite{corps2022chaos}, where it was argued that the parity-breaking asymmetry causes the chaotic domain to develop separately within each of the energy wells. Similarly to the previous case, when $ E\approx -1.5$ and $\alpha=0.7$, we observe two disconnected wells where the left well exhibits chaos and the right well displays regular behavior. These configurations are depicted in Figs. \ref{fig:figure3}(\textbf{c}) and (\textbf{d}), respectively. Note that the available evolution space has increased. As the energy and deformation strength increase to $E\approx-1$ and $\alpha=1.1$, Fig. \ref{fig:figure3}(\textbf{e}) shows a regular trajectory with a toroidal structure evolving in the right well. This observation can be confirmed by tracking the temporal evolution of the variables ($q,p,Q,P$) for theoretical predictions and ($I_{L_1},V_{C_1},I_{L_2},V_{C_2}$) for experiment results.

For the sake of completeness, we examined two parameter configurations near $\alpha_c$: $\alpha=1.41$ [see Fig. \ref{fig:figure3}(\textbf{f})] and $\alpha=1.43$ [see Fig. \ref{fig:figure3}(\textbf{g})]. When $\alpha=1.41\lesssim \alpha_{c}$, the system still displays two disconnected wells, but the potential barrier between them is small due to the proximity of the critical value. For this case, we have prepared an initial condition in the right well to determine if the electronic noise is sufficient to prompt the trajectory to move towards the left well. Our findings suggest that stochastic fluctuations are not strong enough to induce the trajectory to shift towards the left well as illustrated in Fig. \ref{fig:figure3}(\textbf{f}). It is worth noting that at the critical value of deformation strength, $\alpha=\alpha_c$, the right well disappears, causing the energy at which the second well appears and the energy of the logarithmic excited state quantum phase transition to collapse. Under these conditions, the system presents an inflection point with stationary dynamics.  As depicted in Fig. \ref{fig:figure3}(\textbf{g}), for $\alpha=1.43\approx \alpha_{c}$, the trajectory ends in the left well exhibiting regular behavior.

Finally, we have analyzed two different values of deformation strength beyond $\alpha_c$. As mentioned earlier, for $\alpha>\alpha_c$, the system presents only one well. We explored an initial condition whose dynamics is chaotic for $\alpha=1.5$, found at sufficiently high energy, $E\approx 2$, in Fig. \ref{fig:figure3}(\textbf{h}). On the other hand, for $E\approx -12.0001$ and $\alpha=1.5$, the regular dynamic governs the phase plane, as shown in Fig. \ref{fig:figure3}(\textbf{i}). This is because this trajectory has an energy close to the ground-state energy, where the deformed Dicke model can be described by a set of adiabatic invariants that make it approximately integrable \cite{corps2022chaos,Relano2016,Bastarrachea2017}.

\section{CONCLUSIONS}
\label{sec:conclusion}

We have experimentally implemented the semiclassical version of the deformed Dicke model using a modified state-of-art electronic platform that comprises active networks of operational amplifiers and passive components. The parameter configuration via external voltages in our bi-parametric electronic platform enabled the experimental investigation of the rich dynamics of the system and the direct observation of regular and chaotic dynamics in different regions of the space of parameters. Particularly, we have tracked experimentally the ground state as a function of the coupling strength $\gamma$ and deformation strength $\alpha$. In addition, we have outlined a two-dimensional phase diagram of the classical version of the deformed Dicke model as a function of $\gamma$ and $\alpha$ and explored representative phenomena such as the transition from two equivalent wells to a single deformed well, the dynamics of the ground state, as well as the asymmetry of the energy wells. Importantly, the parity-symmetry breaking induced by the deformation strength is clearly unveiled by our experimental measurements. The results of our experiments have been in excellent agreement with the expected results from numerical simulations, demonstrating the
usefulness
of our electronic platform in accurately modeling the semiclassical version of the deformed Dicke model.  


\section*{Acknowledgments}
MAQ-J thankfully acknowledges financial support by CONAHCyT under the Project CF-2023-I-1496 and by DGAPA-UNAM under the Project UNAM-PAPIIT TA101023. J.G.H. received partial financial support from project UNAM-PAPIIT IN109523. A.L.C., R.A.M. and A.R. are supported by the Spanish grants PGC2018-094180-B-I00 and PID2019-106820RB-C21 funded by Ministerio de Ciencia e Innovaci\'{o}n/Agencia Estatal de Investigaci\'{o}n MCIN/AEI/10.13039/501100011033 and FEDER ``A way of making Europe".
A.L.C. acknowledges financial support from `la Caixa' Foundation (ID 100010434) through the fellowship LCF/BQ/DR21/11880024. R.J.L.-M. thankfully acknowledges financial support by DGAPA-UNAM under the project UNAM-PAPIIT IN101623. The Version of Record of this
article is published in The European Physical Journal Plus, and is available online at https://doi.org/10.1140/epjp/s13360-023-04391-6.

\section*{Competing interests}
The authors declare no competing interests.

\section*{Data Availability}
The data that support the findings of this study are available from the corresponding authors upon reasonable request.

\appendix
\section{APPENDIX: EXPERIMENTAL SETUP}

In this Appendix, we describe the experimental implementation of the modified version of the Dicke model. In particular, we have implemented active electrical networks that accurately reproduce the dynamics of the motion equations described by Eq. (\ref{eq:motionpq}). These networks consist of specialized electronic components such as operational amplifiers, resistors, capacitors, and inductors, that are connected together to form circuits that perform a wide range of mathematical operations, from simple addition and subtraction to complex differential equations. Here, the voltages and currents of the circuit represent the continuous physical quantities of the studied system.  We use four configurations of operational amplifiers, namely, adder amplifier, differential amplifier, integrator, and gain amplifier, to construct the linear terms of the system (3). These configurations are constructed with metal resistors (1\% tolerance), polyester capacitors, and general-purpose operational amplifiers LF353. To implement nonlinear functions, such as squaring, square-rooting, multiplication, and division between variables, we have used analog multipliers AD633JN with a typical error of less than 1\%. 

Importantly, we include scaling factors in gain amplifier configurations to avoid the amplitudes of the variables to exceed power supply limits. Particularly, square-rooting operation demands accurate control of the amplitudes because of its high sensitivity to negative signals. Intrinsic parameters such as the intensity of the deformation $(\alpha)$, the coupling strength $(\gamma)$,  the frequency of the quantized radiation field $(\omega)$, and the atomic frequency $(\omega_0)$, as well as the initial conditions $(I_{L_1}(0),V_{C_1}(0),I_{L_2}(0),V_{C_2}(0))$ are set via external voltage signals. These signals are generated by individual digital-to-analog converters which are communicated to a master 8-bit microcontroller (PIC18 family) by the serial peripheral interface (SPI) protocol using the control and data bits, namely, LDAC, DATA, and CLK. Specifically, we use DACs with series MCP492, which offer a resolution of 12 bits, with a reference voltage of 5 volts. Consequently, the voltage resolution is 1.22 mV.  This resolution allows us to set parameter values with up to three significant figures. Accordingly, the initial conditions that have been used for experimentally reproducing results in Fig. \ref{fig:figure3} and are listed in the text and at the end of the figure’s caption have at most three decimals. The initial conditions are set at the output of each integrator amplifier. We have included analog switches to isolate the digital and analog components once the initial conditions are established. Since these devices are switched to high impedance, the dynamics of the system are not affected by the digital circuit. In order to power the device, a bipolar DC power supply is utilized to provide the ±12V bias voltage to both the OPAMPs and the analog multipliers.

\bibliography{reference.bib}

\begin{thebibliography}{63}%
\makeatletter
\providecommand \@ifxundefined [1]{%
 \@ifx{#1\undefined}
}%
\providecommand \@ifnum [1]{%
 \ifnum #1\expandafter \@firstoftwo
 \else \expandafter \@secondoftwo
 \fi
}%
\providecommand \@ifx [1]{%
 \ifx #1\expandafter \@firstoftwo
 \else \expandafter \@secondoftwo
 \fi
}%
\providecommand \natexlab [1]{#1}%
\providecommand \enquote  [1]{``#1''}%
\providecommand \bibnamefont  [1]{#1}%
\providecommand \bibfnamefont [1]{#1}%
\providecommand \citenamefont [1]{#1}%
\providecommand \href@noop [0]{\@secondoftwo}%
\providecommand \href [0]{\begingroup \@sanitize@url \@href}%
\providecommand \@href[1]{\@@startlink{#1}\@@href}%
\providecommand \@@href[1]{\endgroup#1\@@endlink}%
\providecommand \@sanitize@url [0]{\catcode `\\12\catcode `\$12\catcode
  `\&12\catcode `\#12\catcode `\^12\catcode `\_12\catcode `\%12\relax}%
\providecommand \@@startlink[1]{}%
\providecommand \@@endlink[0]{}%
\providecommand \url  [0]{\begingroup\@sanitize@url \@url }%
\providecommand \@url [1]{\endgroup\@href {#1}{\urlprefix }}%
\providecommand \urlprefix  [0]{URL }%
\providecommand \Eprint [0]{\href }%
\providecommand \doibase [0]{https://doi.org/}%
\providecommand \selectlanguage [0]{\@gobble}%
\providecommand \bibinfo  [0]{\@secondoftwo}%
\providecommand \bibfield  [0]{\@secondoftwo}%
\providecommand \translation [1]{[#1]}%
\providecommand \BibitemOpen [0]{}%
\providecommand \bibitemStop [0]{}%
\providecommand \bibitemNoStop [0]{.\EOS\space}%
\providecommand \EOS [0]{\spacefactor3000\relax}%
\providecommand \BibitemShut  [1]{\csname bibitem#1\endcsname}%
\let\auto@bib@innerbib\@empty
\bibitem [{\citenamefont {Baumann}\ \emph {et~al.}(2010)\citenamefont
  {Baumann}, \citenamefont {Guerlin}, \citenamefont {Brennecke},\ and\
  \citenamefont {Esslinger}}]{Baumann2010}%
  \BibitemOpen
  \bibfield  {author} {\bibinfo {author} {\bibfnamefont {K.}~\bibnamefont
  {Baumann}}, \bibinfo {author} {\bibfnamefont {C.}~\bibnamefont {Guerlin}},
  \bibinfo {author} {\bibfnamefont {F.}~\bibnamefont {Brennecke}},\ and\
  \bibinfo {author} {\bibfnamefont {T.}~\bibnamefont {Esslinger}},\ }\href
  {https://doi.org/10.1038/nature09009} {\bibfield  {journal} {\bibinfo
  {journal} {Nature}\ }\textbf {\bibinfo {volume} {464}},\ \bibinfo {pages}
  {1301–1306} (\bibinfo {year} {2010})}\BibitemShut {NoStop}%
\bibitem [{\citenamefont {Muniz}\ \emph {et~al.}(2020)\citenamefont {Muniz},
  \citenamefont {Barberena}, \citenamefont {Lewis-Swan}, \citenamefont {Young},
  \citenamefont {Cline}, \citenamefont {Rey},\ and\ \citenamefont
  {Thompson}}]{Muniz2020}%
  \BibitemOpen
  \bibfield  {author} {\bibinfo {author} {\bibfnamefont {J.}~\bibnamefont
  {Muniz}}, \bibinfo {author} {\bibfnamefont {D.}~\bibnamefont {Barberena}},
  \bibinfo {author} {\bibfnamefont {R.}~\bibnamefont {Lewis-Swan}}, \bibinfo
  {author} {\bibfnamefont {D.}~\bibnamefont {Young}}, \bibinfo {author}
  {\bibfnamefont {J.}~\bibnamefont {Cline}}, \bibinfo {author} {\bibfnamefont
  {A.}~\bibnamefont {Rey}},\ and\ \bibinfo {author} {\bibfnamefont
  {J.}~\bibnamefont {Thompson}},\ }\href
  {https://doi.org/10.1038/s41586-020-2224-x} {\bibfield  {journal} {\bibinfo
  {journal} {Nature}\ }\textbf {\bibinfo {volume} {580}},\ \bibinfo {pages}
  {602} (\bibinfo {year} {2020})}\BibitemShut {NoStop}%
\bibitem [{\citenamefont {Sachdev}(2011)}]{sachdev2011}%
  \BibitemOpen
  \bibfield  {author} {\bibinfo {author} {\bibfnamefont {S.}~\bibnamefont
  {Sachdev}},\ }\href@noop {} {\emph {\bibinfo {title} {Quantum phase
  transitions (2nd edition)}}}\ (\bibinfo  {publisher} {Cambridge University
  Press},\ \bibinfo {year} {2011})\BibitemShut {NoStop}%
\bibitem [{\citenamefont {Cejnar}\ \emph {et~al.}(2021)\citenamefont {Cejnar},
  \citenamefont {Str\'{a}nsk\'{y}}, \citenamefont {Macek},\ and\ \citenamefont
  {Kloc}}]{cejnar2021}%
  \BibitemOpen
  \bibfield  {author} {\bibinfo {author} {\bibfnamefont {P.}~\bibnamefont
  {Cejnar}}, \bibinfo {author} {\bibfnamefont {P.}~\bibnamefont
  {Str\'{a}nsk\'{y}}}, \bibinfo {author} {\bibfnamefont {M.}~\bibnamefont
  {Macek}},\ and\ \bibinfo {author} {\bibfnamefont {M.}~\bibnamefont {Kloc}},\
  }\href {https://doi.org/10.1088/1751-8121/abdfe8} {\bibfield  {journal}
  {\bibinfo  {journal} {Journal of Physics A: Mathematical and Theoretical}\
  }\textbf {\bibinfo {volume} {54}},\ \bibinfo {pages} {133001} (\bibinfo
  {year} {2021})}\BibitemShut {NoStop}%
\bibitem [{\citenamefont {Dicke}(1954)}]{Dicke1954}%
  \BibitemOpen
  \bibfield  {author} {\bibinfo {author} {\bibfnamefont {R.~H.}\ \bibnamefont
  {Dicke}},\ }\href {https://doi.org/10.1103/PhysRev.93.99} {\bibfield
  {journal} {\bibinfo  {journal} {Phys. Rev.}\ }\textbf {\bibinfo {volume}
  {93}},\ \bibinfo {pages} {99} (\bibinfo {year} {1954})}\BibitemShut {NoStop}%
\bibitem [{\citenamefont {Corps}\ and\ \citenamefont
  {Rela\~no}(2022{\natexlab{a}})}]{corps2021PRA}%
  \BibitemOpen
  \bibfield  {author} {\bibinfo {author} {\bibfnamefont {A.~L.}\ \bibnamefont
  {Corps}}\ and\ \bibinfo {author} {\bibfnamefont {A.}~\bibnamefont
  {Rela\~no}},\ }\href {https://doi.org/10.1103/PhysRevA.105.052204} {\bibfield
   {journal} {\bibinfo  {journal} {Phys. Rev. A}\ }\textbf {\bibinfo {volume}
  {105}},\ \bibinfo {pages} {052204} (\bibinfo {year}
  {2022}{\natexlab{a}})}\BibitemShut {NoStop}%
\bibitem [{\citenamefont {Corps}\ and\ \citenamefont
  {Rela\~no}(2021)}]{Corps2021PRL}%
  \BibitemOpen
  \bibfield  {author} {\bibinfo {author} {\bibfnamefont {A.~L.}\ \bibnamefont
  {Corps}}\ and\ \bibinfo {author} {\bibfnamefont {A.}~\bibnamefont
  {Rela\~no}},\ }\href {https://doi.org/10.1103/PhysRevLett.127.130602}
  {\bibfield  {journal} {\bibinfo  {journal} {Phys. Rev. Lett.}\ }\textbf
  {\bibinfo {volume} {127}},\ \bibinfo {pages} {130602} (\bibinfo {year}
  {2021})}\BibitemShut {NoStop}%
\bibitem [{\citenamefont {Bastarrachea-Magnani}\ \emph
  {et~al.}(2014{\natexlab{a}})\citenamefont {Bastarrachea-Magnani},
  \citenamefont {Lerma-Hern\'andez},\ and\ \citenamefont
  {Hirsch}}]{Bastarrachea2014a}%
  \BibitemOpen
  \bibfield  {author} {\bibinfo {author} {\bibfnamefont {M.~A.}\ \bibnamefont
  {Bastarrachea-Magnani}}, \bibinfo {author} {\bibfnamefont {S.}~\bibnamefont
  {Lerma-Hern\'andez}},\ and\ \bibinfo {author} {\bibfnamefont {J.~G.}\
  \bibnamefont {Hirsch}},\ }\href {https://doi.org/10.1103/PhysRevA.89.032101}
  {\bibfield  {journal} {\bibinfo  {journal} {Phys. Rev. A}\ }\textbf {\bibinfo
  {volume} {89}},\ \bibinfo {pages} {032101} (\bibinfo {year}
  {2014}{\natexlab{a}})}\BibitemShut {NoStop}%
\bibitem [{\citenamefont {Puebla}\ \emph {et~al.}(2013)\citenamefont {Puebla},
  \citenamefont {Rela\~no},\ and\ \citenamefont {Retamosa}}]{Puebla2013}%
  \BibitemOpen
  \bibfield  {author} {\bibinfo {author} {\bibfnamefont {R.}~\bibnamefont
  {Puebla}}, \bibinfo {author} {\bibfnamefont {A.}~\bibnamefont {Rela\~no}},\
  and\ \bibinfo {author} {\bibfnamefont {J.}~\bibnamefont {Retamosa}},\ }\href
  {https://doi.org/10.1103/PhysRevA.87.023819} {\bibfield  {journal} {\bibinfo
  {journal} {Phys. Rev. A}\ }\textbf {\bibinfo {volume} {87}},\ \bibinfo
  {pages} {023819} (\bibinfo {year} {2013})}\BibitemShut {NoStop}%
\bibitem [{\citenamefont {Rela\~no}\ \emph {et~al.}(2008)\citenamefont
  {Rela\~no}, \citenamefont {Arias}, \citenamefont {Dukelsky}, \citenamefont
  {Garc\'{\i}a-Ramos},\ and\ \citenamefont {P\'erez-Fern\'andez}}]{Relano2008}%
  \BibitemOpen
  \bibfield  {author} {\bibinfo {author} {\bibfnamefont {A.}~\bibnamefont
  {Rela\~no}}, \bibinfo {author} {\bibfnamefont {J.~M.}\ \bibnamefont {Arias}},
  \bibinfo {author} {\bibfnamefont {J.}~\bibnamefont {Dukelsky}}, \bibinfo
  {author} {\bibfnamefont {J.~E.}\ \bibnamefont {Garc\'{\i}a-Ramos}},\ and\
  \bibinfo {author} {\bibfnamefont {P.}~\bibnamefont {P\'erez-Fern\'andez}},\
  }\href {https://doi.org/10.1103/PhysRevA.78.060102} {\bibfield  {journal}
  {\bibinfo  {journal} {Phys. Rev. A}\ }\textbf {\bibinfo {volume} {78}},\
  \bibinfo {pages} {060102} (\bibinfo {year} {2008})}\BibitemShut {NoStop}%
\bibitem [{\citenamefont {Str\'ansk\'y}\ \emph {et~al.}(2021)\citenamefont
  {Str\'ansk\'y}, \citenamefont {Cejnar},\ and\ \citenamefont
  {Filip}}]{Stransky2021}%
  \BibitemOpen
  \bibfield  {author} {\bibinfo {author} {\bibfnamefont {P.}~\bibnamefont
  {Str\'ansk\'y}}, \bibinfo {author} {\bibfnamefont {P.}~\bibnamefont
  {Cejnar}},\ and\ \bibinfo {author} {\bibfnamefont {R.}~\bibnamefont
  {Filip}},\ }\href {https://doi.org/10.1103/PhysRevA.104.053722} {\bibfield
  {journal} {\bibinfo  {journal} {Phys. Rev. A}\ }\textbf {\bibinfo {volume}
  {104}},\ \bibinfo {pages} {053722} (\bibinfo {year} {2021})}\BibitemShut
  {NoStop}%
\bibitem [{\citenamefont {Hwang}\ \emph {et~al.}(2015)\citenamefont {Hwang},
  \citenamefont {Puebla},\ and\ \citenamefont {Plenio}}]{Hwang2015}%
  \BibitemOpen
  \bibfield  {author} {\bibinfo {author} {\bibfnamefont {M.-J.}\ \bibnamefont
  {Hwang}}, \bibinfo {author} {\bibfnamefont {R.}~\bibnamefont {Puebla}},\ and\
  \bibinfo {author} {\bibfnamefont {M.~B.}\ \bibnamefont {Plenio}},\ }\href
  {https://doi.org/10.1103/PhysRevLett.115.180404} {\bibfield  {journal}
  {\bibinfo  {journal} {Phys. Rev. Lett.}\ }\textbf {\bibinfo {volume} {115}},\
  \bibinfo {pages} {180404} (\bibinfo {year} {2015})}\BibitemShut {NoStop}%
\bibitem [{\citenamefont {Puebla}\ \emph {et~al.}(2016)\citenamefont {Puebla},
  \citenamefont {Hwang},\ and\ \citenamefont {Plenio}}]{Puebla2016}%
  \BibitemOpen
  \bibfield  {author} {\bibinfo {author} {\bibfnamefont {R.}~\bibnamefont
  {Puebla}}, \bibinfo {author} {\bibfnamefont {M.-J.}\ \bibnamefont {Hwang}},\
  and\ \bibinfo {author} {\bibfnamefont {M.~B.}\ \bibnamefont {Plenio}},\
  }\href {https://doi.org/10.1103/PhysRevA.94.023835} {\bibfield  {journal}
  {\bibinfo  {journal} {Phys. Rev. A}\ }\textbf {\bibinfo {volume} {94}},\
  \bibinfo {pages} {023835} (\bibinfo {year} {2016})}\BibitemShut {NoStop}%
\bibitem [{\citenamefont {Klinder}\ \emph {et~al.}(2015)\citenamefont
  {Klinder}, \citenamefont {Ke\ss~ler}, \citenamefont {Wolke}, \citenamefont
  {Mathey},\ and\ \citenamefont {Hemmerich}}]{Klinder2015}%
  \BibitemOpen
  \bibfield  {author} {\bibinfo {author} {\bibfnamefont {J.}~\bibnamefont
  {Klinder}}, \bibinfo {author} {\bibfnamefont {H.}~\bibnamefont {Ke\ss~ler}},
  \bibinfo {author} {\bibfnamefont {M.}~\bibnamefont {Wolke}}, \bibinfo
  {author} {\bibfnamefont {L.}~\bibnamefont {Mathey}},\ and\ \bibinfo {author}
  {\bibfnamefont {A.}~\bibnamefont {Hemmerich}},\ }\href
  {https://doi.org/10.1073/pnas.1417132112} {\bibfield  {journal} {\bibinfo
  {journal} {Proceedings of the National Academy of Sciences}\ }\textbf
  {\bibinfo {volume} {112}},\ \bibinfo {pages} {3290} (\bibinfo {year}
  {2015})}\BibitemShut {NoStop}%
\bibitem [{\citenamefont {Mut-Petit}\ \emph {et~al.}(2018)\citenamefont
  {Mut-Petit}, \citenamefont {Rela\~{n}o}, \citenamefont {Molina},\ and\
  \citenamefont {Jaksch}}]{MurPetit2018}%
  \BibitemOpen
  \bibfield  {author} {\bibinfo {author} {\bibfnamefont {J.}~\bibnamefont
  {Mut-Petit}}, \bibinfo {author} {\bibfnamefont {A.}~\bibnamefont
  {Rela\~{n}o}}, \bibinfo {author} {\bibfnamefont {R.~A.}\ \bibnamefont
  {Molina}},\ and\ \bibinfo {author} {\bibfnamefont {D.}~\bibnamefont
  {Jaksch}},\ }\href {https://doi.org/10.1038/s41467-018-04407-1} {\bibfield
  {journal} {\bibinfo  {journal} {Nature Communications}\ }\textbf {\bibinfo
  {volume} {9}},\ \bibinfo {pages} {2006} (\bibinfo {year} {2018})}\BibitemShut
  {NoStop}%
\bibitem [{\citenamefont {P\'erez-Fern\'andez}\ \emph
  {et~al.}(2011{\natexlab{a}})\citenamefont {P\'erez-Fern\'andez},
  \citenamefont {Rela\~no}, \citenamefont {Arias}, \citenamefont {Cejnar},
  \citenamefont {Dukelsky},\ and\ \citenamefont
  {Garc\'{\i}a-Ramos}}]{Relano2011}%
  \BibitemOpen
  \bibfield  {author} {\bibinfo {author} {\bibfnamefont {P.}~\bibnamefont
  {P\'erez-Fern\'andez}}, \bibinfo {author} {\bibfnamefont {A.}~\bibnamefont
  {Rela\~no}}, \bibinfo {author} {\bibfnamefont {J.~M.}\ \bibnamefont {Arias}},
  \bibinfo {author} {\bibfnamefont {P.}~\bibnamefont {Cejnar}}, \bibinfo
  {author} {\bibfnamefont {J.}~\bibnamefont {Dukelsky}},\ and\ \bibinfo
  {author} {\bibfnamefont {J.~E.}\ \bibnamefont {Garc\'{\i}a-Ramos}},\ }\href
  {https://doi.org/10.1103/PhysRevE.83.046208} {\bibfield  {journal} {\bibinfo
  {journal} {Phys. Rev. E}\ }\textbf {\bibinfo {volume} {83}},\ \bibinfo
  {pages} {046208} (\bibinfo {year} {2011}{\natexlab{a}})}\BibitemShut
  {NoStop}%
\bibitem [{\citenamefont {Bastarrachea-Magnani}\ \emph
  {et~al.}(2014{\natexlab{b}})\citenamefont {Bastarrachea-Magnani},
  \citenamefont {Lerma-Hern\'andez},\ and\ \citenamefont
  {Hirsch}}]{Bastarrachea2014b}%
  \BibitemOpen
  \bibfield  {author} {\bibinfo {author} {\bibfnamefont {M.~A.}\ \bibnamefont
  {Bastarrachea-Magnani}}, \bibinfo {author} {\bibfnamefont {S.}~\bibnamefont
  {Lerma-Hern\'andez}},\ and\ \bibinfo {author} {\bibfnamefont {J.~G.}\
  \bibnamefont {Hirsch}},\ }\href {https://doi.org/10.1103/PhysRevA.89.032102}
  {\bibfield  {journal} {\bibinfo  {journal} {Phys. Rev. A}\ }\textbf {\bibinfo
  {volume} {89}},\ \bibinfo {pages} {032102} (\bibinfo {year}
  {2014}{\natexlab{b}})}\BibitemShut {NoStop}%
\bibitem [{\citenamefont {Lewis-Swan}\ \emph {et~al.}(2019)\citenamefont
  {Lewis-Swan}, \citenamefont {Safavi-Naini}, \citenamefont {Bollinger},\ and\
  \citenamefont {Rey}}]{LewisSwan2019}%
  \BibitemOpen
  \bibfield  {author} {\bibinfo {author} {\bibfnamefont {R.}~\bibnamefont
  {Lewis-Swan}}, \bibinfo {author} {\bibfnamefont {A.}~\bibnamefont
  {Safavi-Naini}}, \bibinfo {author} {\bibfnamefont {J.}~\bibnamefont
  {Bollinger}},\ and\ \bibinfo {author} {\bibfnamefont {A.}~\bibnamefont
  {Rey}},\ }\href {https://doi.org/10.1038/s41467-019-09436-y} {\bibfield
  {journal} {\bibinfo  {journal} {Nature Communications}\ }\textbf {\bibinfo
  {volume} {10}},\ \bibinfo {pages} {1581} (\bibinfo {year}
  {2019})}\BibitemShut {NoStop}%
\bibitem [{\citenamefont {Pilatowsky-Cameo}\ \emph {et~al.}(2021)\citenamefont
  {Pilatowsky-Cameo}, \citenamefont {Villase\~{n}or}, \citenamefont
  {Bastarrachea-Magnani}, \citenamefont {Lerma-Hern\'{a}ndez}, \citenamefont
  {Santos},\ and\ \citenamefont {Hirsch}}]{Cameo2021}%
  \BibitemOpen
  \bibfield  {author} {\bibinfo {author} {\bibfnamefont {S.}~\bibnamefont
  {Pilatowsky-Cameo}}, \bibinfo {author} {\bibfnamefont {D.}~\bibnamefont
  {Villase\~{n}or}}, \bibinfo {author} {\bibfnamefont {M.}~\bibnamefont
  {Bastarrachea-Magnani}}, \bibinfo {author} {\bibfnamefont {S.}~\bibnamefont
  {Lerma-Hern\'{a}ndez}}, \bibinfo {author} {\bibfnamefont {L.}~\bibnamefont
  {Santos}},\ and\ \bibinfo {author} {\bibfnamefont {J.}~\bibnamefont
  {Hirsch}},\ }\href {https://doi.org/doi.org/10.1038/s41467-021-21123-5}
  {\bibfield  {journal} {\bibinfo  {journal} {Nat. Commun.}\ ,\ \bibinfo
  {pages} {865}} (\bibinfo {year} {2021})}\BibitemShut {NoStop}%
\bibitem [{\citenamefont {Pilatowsky-Cameo}\ \emph {et~al.}(2020)\citenamefont
  {Pilatowsky-Cameo}, \citenamefont {Ch\'avez-Carlos}, \citenamefont
  {Bastarrachea-Magnani}, \citenamefont {Str\'ansk\'y}, \citenamefont
  {Lerma-Hern\'andez}, \citenamefont {Santos},\ and\ \citenamefont
  {Hirsch}}]{Cameo2020}%
  \BibitemOpen
  \bibfield  {author} {\bibinfo {author} {\bibfnamefont {S.}~\bibnamefont
  {Pilatowsky-Cameo}}, \bibinfo {author} {\bibfnamefont {J.}~\bibnamefont
  {Ch\'avez-Carlos}}, \bibinfo {author} {\bibfnamefont {M.~A.}\ \bibnamefont
  {Bastarrachea-Magnani}}, \bibinfo {author} {\bibfnamefont {P.}~\bibnamefont
  {Str\'ansk\'y}}, \bibinfo {author} {\bibfnamefont {S.}~\bibnamefont
  {Lerma-Hern\'andez}}, \bibinfo {author} {\bibfnamefont {L.~F.}\ \bibnamefont
  {Santos}},\ and\ \bibinfo {author} {\bibfnamefont {J.~G.}\ \bibnamefont
  {Hirsch}},\ }\href {https://doi.org/10.1103/PhysRevE.101.010202} {\bibfield
  {journal} {\bibinfo  {journal} {Phys. Rev. E}\ }\textbf {\bibinfo {volume}
  {101}},\ \bibinfo {pages} {010202} (\bibinfo {year} {2020})}\BibitemShut
  {NoStop}%
\bibitem [{\citenamefont {Rela\~no}(2018)}]{Relano2018}%
  \BibitemOpen
  \bibfield  {author} {\bibinfo {author} {\bibfnamefont {A.}~\bibnamefont
  {Rela\~no}},\ }\href {https://doi.org/10.1103/PhysRevLett.121.030602}
  {\bibfield  {journal} {\bibinfo  {journal} {Phys. Rev. Lett.}\ }\textbf
  {\bibinfo {volume} {121}},\ \bibinfo {pages} {030602} (\bibinfo {year}
  {2018})}\BibitemShut {NoStop}%
\bibitem [{\citenamefont {Kloc}\ \emph {et~al.}(2018)\citenamefont {Kloc},
  \citenamefont {Str\'ansk\'y},\ and\ \citenamefont {Cejnar}}]{Kloc2018}%
  \BibitemOpen
  \bibfield  {author} {\bibinfo {author} {\bibfnamefont {M.}~\bibnamefont
  {Kloc}}, \bibinfo {author} {\bibfnamefont {P.}~\bibnamefont {Str\'ansk\'y}},\
  and\ \bibinfo {author} {\bibfnamefont {P.}~\bibnamefont {Cejnar}},\ }\href
  {https://doi.org/10.1103/PhysRevA.98.013836} {\bibfield  {journal} {\bibinfo
  {journal} {Phys. Rev. A}\ }\textbf {\bibinfo {volume} {98}},\ \bibinfo
  {pages} {013836} (\bibinfo {year} {2018})}\BibitemShut {NoStop}%
\bibitem [{\citenamefont {Ch\'avez-Carlos}\ \emph {et~al.}(2019)\citenamefont
  {Ch\'avez-Carlos}, \citenamefont {L\'opez-del Carpio}, \citenamefont
  {Bastarrachea-Magnani}, \citenamefont {Str\'ansk\'y}, \citenamefont
  {Lerma-Hern\'andez}, \citenamefont {Santos},\ and\ \citenamefont
  {Hirsch}}]{ChavezCarlos2019}%
  \BibitemOpen
  \bibfield  {author} {\bibinfo {author} {\bibfnamefont {J.}~\bibnamefont
  {Ch\'avez-Carlos}}, \bibinfo {author} {\bibfnamefont {B.}~\bibnamefont
  {L\'opez-del Carpio}}, \bibinfo {author} {\bibfnamefont {M.~A.}\ \bibnamefont
  {Bastarrachea-Magnani}}, \bibinfo {author} {\bibfnamefont {P.}~\bibnamefont
  {Str\'ansk\'y}}, \bibinfo {author} {\bibfnamefont {S.}~\bibnamefont
  {Lerma-Hern\'andez}}, \bibinfo {author} {\bibfnamefont {L.~F.}\ \bibnamefont
  {Santos}},\ and\ \bibinfo {author} {\bibfnamefont {J.~G.}\ \bibnamefont
  {Hirsch}},\ }\href {https://doi.org/10.1103/PhysRevLett.122.024101}
  {\bibfield  {journal} {\bibinfo  {journal} {Phys. Rev. Lett.}\ }\textbf
  {\bibinfo {volume} {122}},\ \bibinfo {pages} {024101} (\bibinfo {year}
  {2019})}\BibitemShut {NoStop}%
\bibitem [{\citenamefont {L\'obez}\ and\ \citenamefont
  {Rela\~no}(2016)}]{Lobez2016}%
  \BibitemOpen
  \bibfield  {author} {\bibinfo {author} {\bibfnamefont {C.~M.}\ \bibnamefont
  {L\'obez}}\ and\ \bibinfo {author} {\bibfnamefont {A.}~\bibnamefont
  {Rela\~no}},\ }\href {https://doi.org/10.1103/PhysRevE.94.012140} {\bibfield
  {journal} {\bibinfo  {journal} {Phys. Rev. E}\ }\textbf {\bibinfo {volume}
  {94}},\ \bibinfo {pages} {012140} (\bibinfo {year} {2016})}\BibitemShut
  {NoStop}%
\bibitem [{\citenamefont {Corps}\ \emph {et~al.}(2022)\citenamefont {Corps},
  \citenamefont {Molina},\ and\ \citenamefont {Rela{\~n}o}}]{corps2022chaos}%
  \BibitemOpen
  \bibfield  {author} {\bibinfo {author} {\bibfnamefont {{\'A}.~L.}\
  \bibnamefont {Corps}}, \bibinfo {author} {\bibfnamefont {R.~A.}\ \bibnamefont
  {Molina}},\ and\ \bibinfo {author} {\bibfnamefont {A.}~\bibnamefont
  {Rela{\~n}o}},\ }\href {https://doi.org/10.1088/1751-8121/ac4b16} {\bibfield
  {journal} {\bibinfo  {journal} {Journal of Physics A: Mathematical and
  Theoretical}\ }\textbf {\bibinfo {volume} {55}},\ \bibinfo {pages} {084001}
  (\bibinfo {year} {2022})}\BibitemShut {NoStop}%
\bibitem [{\citenamefont {Emary}\ and\ \citenamefont
  {Brandes}(2003)}]{Emary2003}%
  \BibitemOpen
  \bibfield  {author} {\bibinfo {author} {\bibfnamefont {C.}~\bibnamefont
  {Emary}}\ and\ \bibinfo {author} {\bibfnamefont {T.}~\bibnamefont
  {Brandes}},\ }\href {https://doi.org/10.1103/PhysRevLett.90.044101}
  {\bibfield  {journal} {\bibinfo  {journal} {Phys. Rev. Lett.}\ }\textbf
  {\bibinfo {volume} {90}},\ \bibinfo {pages} {044101} (\bibinfo {year}
  {2003})}\BibitemShut {NoStop}%
\bibitem [{\citenamefont {Wang}(2022)}]{wang2022}%
  \BibitemOpen
  \bibfield  {author} {\bibinfo {author} {\bibfnamefont {Q.}~\bibnamefont
  {Wang}},\ }\href {https://doi.org/10.3390/e24101415} {\bibfield  {journal}
  {\bibinfo  {journal} {Entropy}\ }\textbf {\bibinfo {volume} {24}},\ \bibinfo
  {pages} {1415} (\bibinfo {year} {2022})}\BibitemShut {NoStop}%
\bibitem [{\citenamefont {Larson}\ and\ \citenamefont
  {Mavrogordatos}(2021)}]{larson2021book}%
  \BibitemOpen
  \bibfield  {author} {\bibinfo {author} {\bibfnamefont {J.}~\bibnamefont
  {Larson}}\ and\ \bibinfo {author} {\bibfnamefont {T.}~\bibnamefont
  {Mavrogordatos}},\ }\href {https://doi.org/10.1088/978-0-7503-3447-1} {\emph
  {\bibinfo {title} {The Jaynes–Cummings Model and Its Descendants, Modern
  research directions}}}\ (\bibinfo  {publisher} {IOP ebooks},\ \bibinfo {year}
  {2021})\BibitemShut {NoStop}%
\bibitem [{\citenamefont {Larson}\ and\ \citenamefont
  {Irish}(2017)}]{larson2017}%
  \BibitemOpen
  \bibfield  {author} {\bibinfo {author} {\bibfnamefont {J.}~\bibnamefont
  {Larson}}\ and\ \bibinfo {author} {\bibfnamefont {E.~K.}\ \bibnamefont
  {Irish}},\ }\href {https://doi.org/10.1088/1751-8121/aa65dc} {\bibfield
  {journal} {\bibinfo  {journal} {Journal of Physics A: Mathematical and
  Theoretical}\ }\textbf {\bibinfo {volume} {50}},\ \bibinfo {pages} {174002}
  (\bibinfo {year} {2017})}\BibitemShut {NoStop}%
\bibitem [{\citenamefont {Corps}\ and\ \citenamefont
  {Rela\~no}(2022{\natexlab{b}})}]{Corpscomment}%
  \BibitemOpen
  \bibfield  {author} {\bibinfo {author} {\bibfnamefont {A.~L.}\ \bibnamefont
  {Corps}}\ and\ \bibinfo {author} {\bibfnamefont {A.}~\bibnamefont
  {Rela\~no}},\ }\href {https://doi.org/10.1103/PhysRevA.106.047701} {\bibfield
   {journal} {\bibinfo  {journal} {Phys. Rev. A}\ }\textbf {\bibinfo {volume}
  {106}},\ \bibinfo {pages} {047701} (\bibinfo {year}
  {2022}{\natexlab{b}})}\BibitemShut {NoStop}%
\bibitem [{\citenamefont {P\'erez-Fern\'andez}\ \emph
  {et~al.}(2011{\natexlab{b}})\citenamefont {P\'erez-Fern\'andez},
  \citenamefont {Cejnar}, \citenamefont {Arias}, \citenamefont {Dukelsky},
  \citenamefont {Garc\'{\i}a-Ramos},\ and\ \citenamefont
  {Rela\~no}}]{PerezFernandez2011}%
  \BibitemOpen
  \bibfield  {author} {\bibinfo {author} {\bibfnamefont {P.}~\bibnamefont
  {P\'erez-Fern\'andez}}, \bibinfo {author} {\bibfnamefont {P.}~\bibnamefont
  {Cejnar}}, \bibinfo {author} {\bibfnamefont {J.~M.}\ \bibnamefont {Arias}},
  \bibinfo {author} {\bibfnamefont {J.}~\bibnamefont {Dukelsky}}, \bibinfo
  {author} {\bibfnamefont {J.~E.}\ \bibnamefont {Garc\'{\i}a-Ramos}},\ and\
  \bibinfo {author} {\bibfnamefont {A.}~\bibnamefont {Rela\~no}},\ }\href
  {https://doi.org/10.1103/PhysRevA.83.033802} {\bibfield  {journal} {\bibinfo
  {journal} {Phys. Rev. A}\ }\textbf {\bibinfo {volume} {83}},\ \bibinfo
  {pages} {033802} (\bibinfo {year} {2011}{\natexlab{b}})}\BibitemShut
  {NoStop}%
\bibitem [{\citenamefont {Brandes}(2013)}]{Brandes2013}%
  \BibitemOpen
  \bibfield  {author} {\bibinfo {author} {\bibfnamefont {T.}~\bibnamefont
  {Brandes}},\ }\href {https://doi.org/10.1103/PhysRevE.88.032133} {\bibfield
  {journal} {\bibinfo  {journal} {Phys. Rev. E}\ }\textbf {\bibinfo {volume}
  {88}},\ \bibinfo {pages} {032133} (\bibinfo {year} {2013})}\BibitemShut
  {NoStop}%
\bibitem [{\citenamefont {De~Aguiar}\ \emph {et~al.}(1992)\citenamefont
  {De~Aguiar}, \citenamefont {Furuya}, \citenamefont {Lewenkopf},\ and\
  \citenamefont {Nemes}}]{de1992chaos}%
  \BibitemOpen
  \bibfield  {author} {\bibinfo {author} {\bibfnamefont {M.}~\bibnamefont
  {De~Aguiar}}, \bibinfo {author} {\bibfnamefont {K.}~\bibnamefont {Furuya}},
  \bibinfo {author} {\bibfnamefont {C.}~\bibnamefont {Lewenkopf}},\ and\
  \bibinfo {author} {\bibfnamefont {M.}~\bibnamefont {Nemes}},\ }\href
  {https://doi.org/10.1016/0003-4916(92)90178-O} {\bibfield  {journal}
  {\bibinfo  {journal} {Annals of Physics}\ }\textbf {\bibinfo {volume}
  {216}},\ \bibinfo {pages} {291} (\bibinfo {year} {1992})}\BibitemShut
  {NoStop}%
\bibitem [{\citenamefont {Ch{\'a}vez-Carlos}\ \emph {et~al.}(2016)\citenamefont
  {Ch{\'a}vez-Carlos}, \citenamefont {Bastarrachea-Magnani}, \citenamefont
  {Lerma-Hern{\'a}ndez},\ and\ \citenamefont {Hirsch}}]{chavez2016classical}%
  \BibitemOpen
  \bibfield  {author} {\bibinfo {author} {\bibfnamefont {J.}~\bibnamefont
  {Ch{\'a}vez-Carlos}}, \bibinfo {author} {\bibfnamefont {M.}~\bibnamefont
  {Bastarrachea-Magnani}}, \bibinfo {author} {\bibfnamefont {S.}~\bibnamefont
  {Lerma-Hern{\'a}ndez}},\ and\ \bibinfo {author} {\bibfnamefont
  {J.}~\bibnamefont {Hirsch}},\ }\href
  {https://doi.org/10.1103/PhysRevE.94.022209} {\bibfield  {journal} {\bibinfo
  {journal} {Physical Review E}\ }\textbf {\bibinfo {volume} {94}},\ \bibinfo
  {pages} {022209} (\bibinfo {year} {2016})}\BibitemShut {NoStop}%
\bibitem [{\citenamefont {Quiroz-Ju{\'a}rez}\ \emph {et~al.}(2020)\citenamefont
  {Quiroz-Ju{\'a}rez}, \citenamefont {Ch{\'a}vez-Carlos}, \citenamefont
  {Arag{\'o}n}, \citenamefont {Hirsch},\ and\ \citenamefont
  {de~J.~Le{\'o}n-Montiel}}]{quiroz2020experimental}%
  \BibitemOpen
  \bibfield  {author} {\bibinfo {author} {\bibfnamefont {M.~A.}\ \bibnamefont
  {Quiroz-Ju{\'a}rez}}, \bibinfo {author} {\bibfnamefont {J.}~\bibnamefont
  {Ch{\'a}vez-Carlos}}, \bibinfo {author} {\bibfnamefont {J.~L.}\ \bibnamefont
  {Arag{\'o}n}}, \bibinfo {author} {\bibfnamefont {J.~G.}\ \bibnamefont
  {Hirsch}},\ and\ \bibinfo {author} {\bibfnamefont {R.}~\bibnamefont
  {de~J.~Le{\'o}n-Montiel}},\ }\href
  {https://doi.org/10.1103/PhysRevResearch.2.033169} {\bibfield  {journal}
  {\bibinfo  {journal} {Physical Review Research}\ }\textbf {\bibinfo {volume}
  {2}},\ \bibinfo {pages} {033169} (\bibinfo {year} {2020})}\BibitemShut
  {NoStop}%
\bibitem [{\citenamefont {Dimer}\ \emph {et~al.}(2007)\citenamefont {Dimer},
  \citenamefont {Estienne}, \citenamefont {Parkins},\ and\ \citenamefont
  {Carmichael}}]{dimer2007proposed}%
  \BibitemOpen
  \bibfield  {author} {\bibinfo {author} {\bibfnamefont {F.}~\bibnamefont
  {Dimer}}, \bibinfo {author} {\bibfnamefont {B.}~\bibnamefont {Estienne}},
  \bibinfo {author} {\bibfnamefont {A.}~\bibnamefont {Parkins}},\ and\ \bibinfo
  {author} {\bibfnamefont {H.}~\bibnamefont {Carmichael}},\ }\href
  {https://doi.org/10.1103/PhysRevA.75.013804} {\bibfield  {journal} {\bibinfo
  {journal} {Physical Review A}\ }\textbf {\bibinfo {volume} {75}},\ \bibinfo
  {pages} {013804} (\bibinfo {year} {2007})}\BibitemShut {NoStop}%
\bibitem [{\citenamefont {Kongkhambut}\ \emph {et~al.}(2021)\citenamefont
  {Kongkhambut}, \citenamefont {Ke{\ss}ler}, \citenamefont {Skulte},
  \citenamefont {Mathey}, \citenamefont {Cosme},\ and\ \citenamefont
  {Hemmerich}}]{kongkhambut2021realization}%
  \BibitemOpen
  \bibfield  {author} {\bibinfo {author} {\bibfnamefont {P.}~\bibnamefont
  {Kongkhambut}}, \bibinfo {author} {\bibfnamefont {H.}~\bibnamefont
  {Ke{\ss}ler}}, \bibinfo {author} {\bibfnamefont {J.}~\bibnamefont {Skulte}},
  \bibinfo {author} {\bibfnamefont {L.}~\bibnamefont {Mathey}}, \bibinfo
  {author} {\bibfnamefont {J.~G.}\ \bibnamefont {Cosme}},\ and\ \bibinfo
  {author} {\bibfnamefont {A.}~\bibnamefont {Hemmerich}},\ }\href
  {https://doi.org/10.1103/PhysRevLett.127.253601} {\bibfield  {journal}
  {\bibinfo  {journal} {Physical Review Letters}\ }\textbf {\bibinfo {volume}
  {127}},\ \bibinfo {pages} {253601} (\bibinfo {year} {2021})}\BibitemShut
  {NoStop}%
\bibitem [{\citenamefont {Gonz{\'a}lez-Tudela}\ and\ \citenamefont
  {Porras}(2013)}]{gonzalez2013mesoscopic}%
  \BibitemOpen
  \bibfield  {author} {\bibinfo {author} {\bibfnamefont {A.}~\bibnamefont
  {Gonz{\'a}lez-Tudela}}\ and\ \bibinfo {author} {\bibfnamefont
  {D.}~\bibnamefont {Porras}},\ }\href
  {https://doi.org/10.1103/PhysRevLett.110.080502} {\bibfield  {journal}
  {\bibinfo  {journal} {Physical Review Letters}\ }\textbf {\bibinfo {volume}
  {110}},\ \bibinfo {pages} {080502} (\bibinfo {year} {2013})}\BibitemShut
  {NoStop}%
\bibitem [{\citenamefont {Mlynek}\ \emph {et~al.}(2014)\citenamefont {Mlynek},
  \citenamefont {Abdumalikov}, \citenamefont {Eichler},\ and\ \citenamefont
  {Wallraff}}]{mlynek2014observation}%
  \BibitemOpen
  \bibfield  {author} {\bibinfo {author} {\bibfnamefont {J.~A.}\ \bibnamefont
  {Mlynek}}, \bibinfo {author} {\bibfnamefont {A.~A.}\ \bibnamefont
  {Abdumalikov}}, \bibinfo {author} {\bibfnamefont {C.}~\bibnamefont
  {Eichler}},\ and\ \bibinfo {author} {\bibfnamefont {A.}~\bibnamefont
  {Wallraff}},\ }\href {https://doi.org/10.1038/ncomms6186} {\bibfield
  {journal} {\bibinfo  {journal} {Nature Communications}\ }\textbf {\bibinfo
  {volume} {5}},\ \bibinfo {pages} {1} (\bibinfo {year} {2014})}\BibitemShut
  {NoStop}%
\bibitem [{\citenamefont {Zhiqiang}\ \emph {et~al.}(2017)\citenamefont
  {Zhiqiang}, \citenamefont {Lee}, \citenamefont {Kumar}, \citenamefont
  {Arnold}, \citenamefont {Masson}, \citenamefont {Parkins},\ and\
  \citenamefont {Barrett}}]{zhiqiang2017nonequilibrium}%
  \BibitemOpen
  \bibfield  {author} {\bibinfo {author} {\bibfnamefont {Z.}~\bibnamefont
  {Zhiqiang}}, \bibinfo {author} {\bibfnamefont {C.~H.}\ \bibnamefont {Lee}},
  \bibinfo {author} {\bibfnamefont {R.}~\bibnamefont {Kumar}}, \bibinfo
  {author} {\bibfnamefont {K.}~\bibnamefont {Arnold}}, \bibinfo {author}
  {\bibfnamefont {S.~J.}\ \bibnamefont {Masson}}, \bibinfo {author}
  {\bibfnamefont {A.}~\bibnamefont {Parkins}},\ and\ \bibinfo {author}
  {\bibfnamefont {M.}~\bibnamefont {Barrett}},\ }\href
  {https://doi.org/10.1364/OPTICA.4.000424} {\bibfield  {journal} {\bibinfo
  {journal} {Optica}\ }\textbf {\bibinfo {volume} {4}},\ \bibinfo {pages} {424}
  (\bibinfo {year} {2017})}\BibitemShut {NoStop}%
\bibitem [{\citenamefont {Masson}\ \emph {et~al.}(2017)\citenamefont {Masson},
  \citenamefont {Barrett},\ and\ \citenamefont {Parkins}}]{masson2017cavity}%
  \BibitemOpen
  \bibfield  {author} {\bibinfo {author} {\bibfnamefont {S.~J.}\ \bibnamefont
  {Masson}}, \bibinfo {author} {\bibfnamefont {M.}~\bibnamefont {Barrett}},\
  and\ \bibinfo {author} {\bibfnamefont {S.}~\bibnamefont {Parkins}},\ }\href
  {https://doi.org/10.1103/PhysRevLett.119.213601} {\bibfield  {journal}
  {\bibinfo  {journal} {Physical Review Letters}\ }\textbf {\bibinfo {volume}
  {119}},\ \bibinfo {pages} {213601} (\bibinfo {year} {2017})}\BibitemShut
  {NoStop}%
\bibitem [{\citenamefont {Carlson}\ \emph {et~al.}(1967)\citenamefont
  {Carlson}, \citenamefont {Hannauer}, \citenamefont {Carey},\ and\
  \citenamefont {Holsberg}}]{carlson1967handbook}%
  \BibitemOpen
  \bibfield  {author} {\bibinfo {author} {\bibfnamefont {A.}~\bibnamefont
  {Carlson}}, \bibinfo {author} {\bibfnamefont {G.}~\bibnamefont {Hannauer}},
  \bibinfo {author} {\bibfnamefont {T.}~\bibnamefont {Carey}},\ and\ \bibinfo
  {author} {\bibfnamefont {P.~J.}\ \bibnamefont {Holsberg}},\ }\href@noop {}
  {\bibfield  {journal} {\bibinfo  {journal} {Inc.: Princeton, New Jersey}\ }
  (\bibinfo {year} {1967})}\BibitemShut {NoStop}%
\bibitem [{\citenamefont {Johnson}(1963)}]{johnson1963analog}%
  \BibitemOpen
  \bibfield  {author} {\bibinfo {author} {\bibfnamefont {C.~L.}\ \bibnamefont
  {Johnson}},\ }\href@noop {} {\emph {\bibinfo {title} {Analog computer
  techniques}}}\ (\bibinfo  {publisher} {McGraw-Hill},\ \bibinfo {year}
  {1963})\BibitemShut {NoStop}%
\bibitem [{\citenamefont {Noordergraaf}\ \emph {et~al.}(1963)\citenamefont
  {Noordergraaf}, \citenamefont {Verdouw},\ and\ \citenamefont
  {Boom}}]{noordergraaf1963use}%
  \BibitemOpen
  \bibfield  {author} {\bibinfo {author} {\bibfnamefont {A.}~\bibnamefont
  {Noordergraaf}}, \bibinfo {author} {\bibfnamefont {P.~D.}\ \bibnamefont
  {Verdouw}},\ and\ \bibinfo {author} {\bibfnamefont {H.~B.}\ \bibnamefont
  {Boom}},\ }\href {https://doi.org/10.1016/S0033-0620(63)80009-2} {\bibfield
  {journal} {\bibinfo  {journal} {Progress in Cardiovascular Diseases}\
  }\textbf {\bibinfo {volume} {5}},\ \bibinfo {pages} {419} (\bibinfo {year}
  {1963})}\BibitemShut {NoStop}%
\bibitem [{\citenamefont {Moore}(1998)}]{moore1998dynamical}%
  \BibitemOpen
  \bibfield  {author} {\bibinfo {author} {\bibfnamefont {C.}~\bibnamefont
  {Moore}},\ }\href {https://doi.org/10.1016/S0304-3975(97)00028-5} {\bibfield
  {journal} {\bibinfo  {journal} {Theoretical Computer Science}\ }\textbf
  {\bibinfo {volume} {201}},\ \bibinfo {pages} {99} (\bibinfo {year}
  {1998})}\BibitemShut {NoStop}%
\bibitem [{\citenamefont {Vazquez-Medina}\ \emph {et~al.}(2013)\citenamefont
  {Vazquez-Medina}, \citenamefont {Jimenez-Ramirez}, \citenamefont
  {Quiroz-Juarez},\ and\ \citenamefont {Aragon}}]{vazquez2013arbitrary}%
  \BibitemOpen
  \bibfield  {author} {\bibinfo {author} {\bibfnamefont {R.}~\bibnamefont
  {Vazquez-Medina}}, \bibinfo {author} {\bibfnamefont {O.}~\bibnamefont
  {Jimenez-Ramirez}}, \bibinfo {author} {\bibfnamefont {M.}~\bibnamefont
  {Quiroz-Juarez}},\ and\ \bibinfo {author} {\bibfnamefont {J.}~\bibnamefont
  {Aragon}},\ }\href {https://doi.org/10.1016/j.chaos.2013.03.006} {\bibfield
  {journal} {\bibinfo  {journal} {Chaos, Solitons \& Fractals}\ }\textbf
  {\bibinfo {volume} {51}},\ \bibinfo {pages} {36} (\bibinfo {year}
  {2013})}\BibitemShut {NoStop}%
\bibitem [{\citenamefont {Quiroz-Ju{\'a}rez}\ \emph {et~al.}(2019)\citenamefont
  {Quiroz-Ju{\'a}rez}, \citenamefont {Jim{\'e}nez-Ram{\'\i}rez}, \citenamefont
  {V{\'a}zquez-Medina}, \citenamefont {Bre{\~n}a-Medina}, \citenamefont
  {Arag{\'o}n},\ and\ \citenamefont {Barrio}}]{quiroz2019generation}%
  \BibitemOpen
  \bibfield  {author} {\bibinfo {author} {\bibfnamefont {M.}~\bibnamefont
  {Quiroz-Ju{\'a}rez}}, \bibinfo {author} {\bibfnamefont {O.}~\bibnamefont
  {Jim{\'e}nez-Ram{\'\i}rez}}, \bibinfo {author} {\bibfnamefont
  {R.}~\bibnamefont {V{\'a}zquez-Medina}}, \bibinfo {author} {\bibfnamefont
  {V.}~\bibnamefont {Bre{\~n}a-Medina}}, \bibinfo {author} {\bibfnamefont
  {J.}~\bibnamefont {Arag{\'o}n}},\ and\ \bibinfo {author} {\bibfnamefont
  {R.}~\bibnamefont {Barrio}},\ }\href
  {https://doi.org/10.1038/s41598-019-55448-5} {\bibfield  {journal} {\bibinfo
  {journal} {Scientific Reports}\ }\textbf {\bibinfo {volume} {9}},\ \bibinfo
  {pages} {1} (\bibinfo {year} {2019})}\BibitemShut {NoStop}%
\bibitem [{\citenamefont {Jim{\'e}nez-Ram{\'\i}rez}\ \emph
  {et~al.}(2021)\citenamefont {Jim{\'e}nez-Ram{\'\i}rez}, \citenamefont
  {Cruz-Dom{\'\i}nguez}, \citenamefont {Quiroz-Ju{\'a}rez}, \citenamefont
  {Aragon},\ and\ \citenamefont
  {V{\'a}zquez-Medina}}]{jimenez2021experimental}%
  \BibitemOpen
  \bibfield  {author} {\bibinfo {author} {\bibfnamefont {O.}~\bibnamefont
  {Jim{\'e}nez-Ram{\'\i}rez}}, \bibinfo {author} {\bibfnamefont
  {E.}~\bibnamefont {Cruz-Dom{\'\i}nguez}}, \bibinfo {author} {\bibfnamefont
  {M.}~\bibnamefont {Quiroz-Ju{\'a}rez}}, \bibinfo {author} {\bibfnamefont
  {J.}~\bibnamefont {Aragon}},\ and\ \bibinfo {author} {\bibfnamefont
  {R.}~\bibnamefont {V{\'a}zquez-Medina}},\ }\href
  {https://doi.org/10.1016/j.csfx.2021.100058} {\bibfield  {journal} {\bibinfo
  {journal} {Chaos, Solitons \& Fractals: X}\ }\textbf {\bibinfo {volume}
  {6}},\ \bibinfo {pages} {100058} (\bibinfo {year} {2021})}\BibitemShut
  {NoStop}%
\bibitem [{\citenamefont {Escobar-Ruiz}\ \emph {et~al.}(2022)\citenamefont
  {Escobar-Ruiz}, \citenamefont {Quiroz-Juarez}, \citenamefont {Del
  Rio-Correa},\ and\ \citenamefont {Aquino}}]{escobar2022classical}%
  \BibitemOpen
  \bibfield  {author} {\bibinfo {author} {\bibfnamefont {A.}~\bibnamefont
  {Escobar-Ruiz}}, \bibinfo {author} {\bibfnamefont {M.}~\bibnamefont
  {Quiroz-Juarez}}, \bibinfo {author} {\bibfnamefont {J.}~\bibnamefont {Del
  Rio-Correa}},\ and\ \bibinfo {author} {\bibfnamefont {N.}~\bibnamefont
  {Aquino}},\ }\href {https://doi.org/10.1038/s41598-022-17541-0} {\bibfield
  {journal} {\bibinfo  {journal} {Scientific Reports}\ }\textbf {\bibinfo
  {volume} {12}},\ \bibinfo {pages} {13346} (\bibinfo {year}
  {2022})}\BibitemShut {NoStop}%
\bibitem [{\citenamefont {Chen}\ \emph {et~al.}(2020)\citenamefont {Chen},
  \citenamefont {Jahanshahi}, \citenamefont {Abba}, \citenamefont
  {Sol{\'\i}s-P{\'e}rez}, \citenamefont {Bekiros}, \citenamefont
  {G{\'o}mez-Aguilar}, \citenamefont {Yousefpour},\ and\ \citenamefont
  {Chu}}]{chen2020effect}%
  \BibitemOpen
  \bibfield  {author} {\bibinfo {author} {\bibfnamefont {S.-B.}\ \bibnamefont
  {Chen}}, \bibinfo {author} {\bibfnamefont {H.}~\bibnamefont {Jahanshahi}},
  \bibinfo {author} {\bibfnamefont {O.~A.}\ \bibnamefont {Abba}}, \bibinfo
  {author} {\bibfnamefont {J.}~\bibnamefont {Sol{\'\i}s-P{\'e}rez}}, \bibinfo
  {author} {\bibfnamefont {S.}~\bibnamefont {Bekiros}}, \bibinfo {author}
  {\bibfnamefont {J.}~\bibnamefont {G{\'o}mez-Aguilar}}, \bibinfo {author}
  {\bibfnamefont {A.}~\bibnamefont {Yousefpour}},\ and\ \bibinfo {author}
  {\bibfnamefont {Y.-M.}\ \bibnamefont {Chu}},\ }\href
  {https://doi.org/10.1016/j.chaos.2020.110223} {\bibfield  {journal} {\bibinfo
   {journal} {Chaos, Solitons \& Fractals}\ }\textbf {\bibinfo {volume}
  {140}},\ \bibinfo {pages} {110223} (\bibinfo {year} {2020})}\BibitemShut
  {NoStop}%
\bibitem [{\citenamefont {de~J.~Le\'on-Montiel}\ \emph
  {et~al.}(2015)\citenamefont {de~J.~Le\'on-Montiel}, \citenamefont
  {Quiroz-Ju{\'a}rez}, \citenamefont {Quintero-Torres}, \citenamefont
  {Dom{\'\i}nguez-Ju{\'a}rez}, \citenamefont {Moya-Cessa}, \citenamefont
  {Torres},\ and\ \citenamefont {Arag{\'o}n}}]{leon2015noise}%
  \BibitemOpen
  \bibfield  {author} {\bibinfo {author} {\bibfnamefont {R.}~\bibnamefont
  {de~J.~Le\'on-Montiel}}, \bibinfo {author} {\bibfnamefont {M.~A.}\
  \bibnamefont {Quiroz-Ju{\'a}rez}}, \bibinfo {author} {\bibfnamefont
  {R.}~\bibnamefont {Quintero-Torres}}, \bibinfo {author} {\bibfnamefont
  {J.~L.}\ \bibnamefont {Dom{\'\i}nguez-Ju{\'a}rez}}, \bibinfo {author}
  {\bibfnamefont {H.~M.}\ \bibnamefont {Moya-Cessa}}, \bibinfo {author}
  {\bibfnamefont {J.~P.}\ \bibnamefont {Torres}},\ and\ \bibinfo {author}
  {\bibfnamefont {J.~L.}\ \bibnamefont {Arag{\'o}n}},\ }\href
  {https://doi.org/10.1038/srep17339} {\bibfield  {journal} {\bibinfo
  {journal} {Scientific Reports}\ }\textbf {\bibinfo {volume} {5}},\ \bibinfo
  {pages} {1} (\bibinfo {year} {2015})}\BibitemShut {NoStop}%
\bibitem [{\citenamefont {Quiroz-Ju{\'a}rez}\ \emph {et~al.}(2017)\citenamefont
  {Quiroz-Ju{\'a}rez}, \citenamefont {Arag{\'o}n}, \citenamefont
  {de~J.~Le\'on-Montiel}, \citenamefont {V{\'a}zquez-Medina}, \citenamefont
  {Dom{\'\i}nguez-Ju{\'a}rez},\ and\ \citenamefont
  {Quintero-Torres}}]{quiroz2017emergence}%
  \BibitemOpen
  \bibfield  {author} {\bibinfo {author} {\bibfnamefont {M.}~\bibnamefont
  {Quiroz-Ju{\'a}rez}}, \bibinfo {author} {\bibfnamefont {J.}~\bibnamefont
  {Arag{\'o}n}}, \bibinfo {author} {\bibfnamefont {R.}~\bibnamefont
  {de~J.~Le\'on-Montiel}}, \bibinfo {author} {\bibfnamefont {R.}~\bibnamefont
  {V{\'a}zquez-Medina}}, \bibinfo {author} {\bibfnamefont {J.}~\bibnamefont
  {Dom{\'\i}nguez-Ju{\'a}rez}},\ and\ \bibinfo {author} {\bibfnamefont
  {R.}~\bibnamefont {Quintero-Torres}},\ }\href
  {https://doi.org/10.1209/0295-5075/116/50004} {\bibfield  {journal} {\bibinfo
   {journal} {EPL (Europhysics Letters)}\ }\textbf {\bibinfo {volume} {116}},\
  \bibinfo {pages} {50004} (\bibinfo {year} {2017})}\BibitemShut {NoStop}%
\bibitem [{\citenamefont {de~J.~Le\'on-Montiel}\ \emph
  {et~al.}(2018)\citenamefont {de~J.~Le\'on-Montiel}, \citenamefont
  {Quiroz-Ju{\'a}rez}, \citenamefont {Dom{\'\i}nguez-Ju{\'a}rez}, \citenamefont
  {Quintero-Torres}, \citenamefont {Arag{\'o}n}, \citenamefont {Harter},\ and\
  \citenamefont {Joglekar}}]{leon2018observation}%
  \BibitemOpen
  \bibfield  {author} {\bibinfo {author} {\bibfnamefont {R.}~\bibnamefont
  {de~J.~Le\'on-Montiel}}, \bibinfo {author} {\bibfnamefont {M.~A.}\
  \bibnamefont {Quiroz-Ju{\'a}rez}}, \bibinfo {author} {\bibfnamefont {J.~L.}\
  \bibnamefont {Dom{\'\i}nguez-Ju{\'a}rez}}, \bibinfo {author} {\bibfnamefont
  {R.}~\bibnamefont {Quintero-Torres}}, \bibinfo {author} {\bibfnamefont
  {J.~L.}\ \bibnamefont {Arag{\'o}n}}, \bibinfo {author} {\bibfnamefont
  {A.~K.}\ \bibnamefont {Harter}},\ and\ \bibinfo {author} {\bibfnamefont
  {Y.~N.}\ \bibnamefont {Joglekar}},\ }\href
  {https://doi.org/10.1038/s42005-018-0087-3} {\bibfield  {journal} {\bibinfo
  {journal} {Communications Physics}\ }\textbf {\bibinfo {volume} {1}},\
  \bibinfo {pages} {1} (\bibinfo {year} {2018})}\BibitemShut {NoStop}%
\bibitem [{\citenamefont {Quiroz-Ju{\'a}rez}\ \emph {et~al.}(2021)\citenamefont
  {Quiroz-Ju{\'a}rez}, \citenamefont {You}, \citenamefont
  {Carrillo-Mart{\'\i}nez}, \citenamefont {Montiel-{\'A}lvarez}, \citenamefont
  {Arag{\'o}n}, \citenamefont {Maga{\~n}a-Loaiza},\ and\ \citenamefont
  {de~J.~Le{\'o}n-Montiel}}]{quiroz2021reconfigurable}%
  \BibitemOpen
  \bibfield  {author} {\bibinfo {author} {\bibfnamefont {M.~A.}\ \bibnamefont
  {Quiroz-Ju{\'a}rez}}, \bibinfo {author} {\bibfnamefont {C.}~\bibnamefont
  {You}}, \bibinfo {author} {\bibfnamefont {J.}~\bibnamefont
  {Carrillo-Mart{\'\i}nez}}, \bibinfo {author} {\bibfnamefont {D.}~\bibnamefont
  {Montiel-{\'A}lvarez}}, \bibinfo {author} {\bibfnamefont {J.~L.}\
  \bibnamefont {Arag{\'o}n}}, \bibinfo {author} {\bibfnamefont {O.~S.}\
  \bibnamefont {Maga{\~n}a-Loaiza}},\ and\ \bibinfo {author} {\bibfnamefont
  {R.}~\bibnamefont {de~J.~Le{\'o}n-Montiel}},\ }\href
  {https://doi.org/10.1103/PhysRevResearch.3.013010} {\bibfield  {journal}
  {\bibinfo  {journal} {Physical Review Research}\ }\textbf {\bibinfo {volume}
  {3}},\ \bibinfo {pages} {013010} (\bibinfo {year} {2021})}\BibitemShut
  {NoStop}%
\bibitem [{\citenamefont {Quiroz-Ju\'arez}\ \emph {et~al.}(2022)\citenamefont
  {Quiroz-Ju\'arez}, \citenamefont {Agarwal}, \citenamefont {Cochran},
  \citenamefont {Arag\'on}, \citenamefont {Joglekar},\ and\ \citenamefont
  {de~J.~Le\'on-Montiel}}]{quiroz2021demand}%
  \BibitemOpen
  \bibfield  {author} {\bibinfo {author} {\bibfnamefont {M.~A.}\ \bibnamefont
  {Quiroz-Ju\'arez}}, \bibinfo {author} {\bibfnamefont {K.~S.}\ \bibnamefont
  {Agarwal}}, \bibinfo {author} {\bibfnamefont {Z.~A.}\ \bibnamefont
  {Cochran}}, \bibinfo {author} {\bibfnamefont {J.~L.}\ \bibnamefont
  {Arag\'on}}, \bibinfo {author} {\bibfnamefont {Y.~N.}\ \bibnamefont
  {Joglekar}},\ and\ \bibinfo {author} {\bibfnamefont {R.}~\bibnamefont
  {de~J.~Le\'on-Montiel}},\ }\href
  {https://doi.org/10.1103/PhysRevApplied.18.054034} {\bibfield  {journal}
  {\bibinfo  {journal} {Phys. Rev. Appl.}\ }\textbf {\bibinfo {volume} {18}},\
  \bibinfo {pages} {054034} (\bibinfo {year} {2022})}\BibitemShut {NoStop}%
\bibitem [{\citenamefont {Dong}\ \emph {et~al.}(2021)\citenamefont {Dong},
  \citenamefont {Juri\ifmmode \check{c}\else \v{c}\fi{}i\ifmmode~\acute{c}\else
  \'{c}\fi{}},\ and\ \citenamefont {Roy}}]{Dong2021}%
  \BibitemOpen
  \bibfield  {author} {\bibinfo {author} {\bibfnamefont {J.}~\bibnamefont
  {Dong}}, \bibinfo {author} {\bibfnamefont {V.}~\bibnamefont {Juri\ifmmode
  \check{c}\else \v{c}\fi{}i\ifmmode~\acute{c}\else \'{c}\fi{}}},\ and\
  \bibinfo {author} {\bibfnamefont {B.}~\bibnamefont {Roy}},\ }\href
  {https://doi.org/10.1103/PhysRevResearch.3.023056} {\bibfield  {journal}
  {\bibinfo  {journal} {Phys. Rev. Res.}\ }\textbf {\bibinfo {volume} {3}},\
  \bibinfo {pages} {023056} (\bibinfo {year} {2021})}\BibitemShut {NoStop}%
\bibitem [{\citenamefont {Hua~Lee}\ \emph {et~al.}(2018)\citenamefont
  {Hua~Lee}, \citenamefont {Imhof}, \citenamefont {Berger}, \citenamefont
  {Bayer}, \citenamefont {Brehm}, \citenamefont {W.~Molenkamp}, \citenamefont
  {Kiessling},\ and\ \citenamefont {Thomale}}]{Hua2018}%
  \BibitemOpen
  \bibfield  {author} {\bibinfo {author} {\bibfnamefont {C.}~\bibnamefont
  {Hua~Lee}}, \bibinfo {author} {\bibfnamefont {S.}~\bibnamefont {Imhof}},
  \bibinfo {author} {\bibfnamefont {C.}~\bibnamefont {Berger}}, \bibinfo
  {author} {\bibfnamefont {F.}~\bibnamefont {Bayer}}, \bibinfo {author}
  {\bibfnamefont {J.}~\bibnamefont {Brehm}}, \bibinfo {author} {\bibfnamefont
  {L.}~\bibnamefont {W.~Molenkamp}}, \bibinfo {author} {\bibfnamefont
  {T.}~\bibnamefont {Kiessling}},\ and\ \bibinfo {author} {\bibfnamefont
  {R.}~\bibnamefont {Thomale}},\ }\href
  {https://doi.org/10.1038/s42005-018-0035-2} {\bibfield  {journal} {\bibinfo
  {journal} {Communications Physics}\ }\textbf {\bibinfo {volume} {1}},\
  \bibinfo {pages} {39} (\bibinfo {year} {2018})}\BibitemShut {NoStop}%
\bibitem [{\citenamefont {Imhof}\ \emph {et~al.}(2018)\citenamefont {Imhof},
  \citenamefont {Berger}, \citenamefont {Bayer}, \citenamefont {Brehm},
  \citenamefont {Molenkamp}, \citenamefont {Kiessling}, \citenamefont
  {Schindler}, \citenamefont {Lee}, \citenamefont {Greiter}, \citenamefont
  {Neupert} \emph {et~al.}}]{imhof2018topolectrical}%
  \BibitemOpen
  \bibfield  {author} {\bibinfo {author} {\bibfnamefont {S.}~\bibnamefont
  {Imhof}}, \bibinfo {author} {\bibfnamefont {C.}~\bibnamefont {Berger}},
  \bibinfo {author} {\bibfnamefont {F.}~\bibnamefont {Bayer}}, \bibinfo
  {author} {\bibfnamefont {J.}~\bibnamefont {Brehm}}, \bibinfo {author}
  {\bibfnamefont {L.~W.}\ \bibnamefont {Molenkamp}}, \bibinfo {author}
  {\bibfnamefont {T.}~\bibnamefont {Kiessling}}, \bibinfo {author}
  {\bibfnamefont {F.}~\bibnamefont {Schindler}}, \bibinfo {author}
  {\bibfnamefont {C.~H.}\ \bibnamefont {Lee}}, \bibinfo {author} {\bibfnamefont
  {M.}~\bibnamefont {Greiter}}, \bibinfo {author} {\bibfnamefont
  {T.}~\bibnamefont {Neupert}}, \emph {et~al.},\ }\href
  {https://doi.org/10.1038/s41567-018-0246-1} {\bibfield  {journal} {\bibinfo
  {journal} {Nature Physics}\ }\textbf {\bibinfo {volume} {14}},\ \bibinfo
  {pages} {925} (\bibinfo {year} {2018})}\BibitemShut {NoStop}%
\bibitem [{\citenamefont {Albert}\ \emph {et~al.}(2015)\citenamefont {Albert},
  \citenamefont {Glazman},\ and\ \citenamefont
  {Jiang}}]{albert2015topological}%
  \BibitemOpen
  \bibfield  {author} {\bibinfo {author} {\bibfnamefont {V.~V.}\ \bibnamefont
  {Albert}}, \bibinfo {author} {\bibfnamefont {L.~I.}\ \bibnamefont
  {Glazman}},\ and\ \bibinfo {author} {\bibfnamefont {L.}~\bibnamefont
  {Jiang}},\ }\href {https://doi.org/10.1103/PhysRevLett.114.173902} {\bibfield
   {journal} {\bibinfo  {journal} {Physical review letters}\ }\textbf {\bibinfo
  {volume} {114}},\ \bibinfo {pages} {173902} (\bibinfo {year}
  {2015})}\BibitemShut {NoStop}%
\bibitem [{\citenamefont {Zhong}\ and\ \citenamefont
  {Ayrom}(1985)}]{Zhong1985}%
  \BibitemOpen
  \bibfield  {author} {\bibinfo {author} {\bibfnamefont {G.-Q.}\ \bibnamefont
  {Zhong}}\ and\ \bibinfo {author} {\bibfnamefont {F.}~\bibnamefont {Ayrom}},\
  }\href {https://doi.org/10.1002/cta.4490130109.} {\bibfield  {journal}
  {\bibinfo  {journal} {International Journal of Circuit Theory and
  Applications}\ }\textbf {\bibinfo {volume} {13}},\ \bibinfo {pages} {93}
  (\bibinfo {year} {1985})}\BibitemShut {NoStop}%
\bibitem [{\citenamefont {Kuznetsov}\ \emph {et~al.}(2023)\citenamefont
  {Kuznetsov}, \citenamefont {Mokaev}, \citenamefont {Ponomarenko},
  \citenamefont {Seleznev}, \citenamefont {Stankevich},\ and\ \citenamefont
  {Chua}}]{Kuznetsov2023}%
  \BibitemOpen
  \bibfield  {author} {\bibinfo {author} {\bibfnamefont {N.}~\bibnamefont
  {Kuznetsov}}, \bibinfo {author} {\bibfnamefont {T.}~\bibnamefont {Mokaev}},
  \bibinfo {author} {\bibfnamefont {V.}~\bibnamefont {Ponomarenko}}, \bibinfo
  {author} {\bibfnamefont {E.}~\bibnamefont {Seleznev}}, \bibinfo {author}
  {\bibfnamefont {N.}~\bibnamefont {Stankevich}},\ and\ \bibinfo {author}
  {\bibfnamefont {L.}~\bibnamefont {Chua}},\ }\href
  {https://doi.org/10.1007/s11071-022-08078-y} {\bibfield  {journal} {\bibinfo
  {journal} {Nonlinear dynamics}\ }\textbf {\bibinfo {volume} {111}},\ \bibinfo
  {pages} {5859} (\bibinfo {year} {2023})}\BibitemShut {NoStop}%
\bibitem [{\citenamefont {Rela{\~n}o}\ \emph {et~al.}(2016)\citenamefont
  {Rela{\~n}o}, \citenamefont {Bastarrachea-Magnani},\ and\ \citenamefont
  {Lerma-Hern\'{a}ndez}}]{Relano2016}%
  \BibitemOpen
  \bibfield  {author} {\bibinfo {author} {\bibfnamefont {A.}~\bibnamefont
  {Rela{\~n}o}}, \bibinfo {author} {\bibfnamefont {M.}~\bibnamefont
  {Bastarrachea-Magnani}},\ and\ \bibinfo {author} {\bibfnamefont
  {S.}~\bibnamefont {Lerma-Hern\'{a}ndez}},\ }\href
  {https://doi.org/10.1209/0295-5075/116/50005} {\bibfield  {journal} {\bibinfo
   {journal} {EPL (Europhysics Letters)}\ }\textbf {\bibinfo {volume} {16}},\
  \bibinfo {pages} {50005} (\bibinfo {year} {2016})}\BibitemShut {NoStop}%
\bibitem [{\citenamefont {Bastarrachea-Magnani}\ \emph
  {et~al.}(2017)\citenamefont {Bastarrachea-Magnani}, \citenamefont
  {Rela{\~n}o}, \citenamefont {Lerma-Hern\'{a}ndez}, \citenamefont
  {L\'{o}pez-del Carpio}, \citenamefont {Ch\'{a}vez-Carlos},\ and\
  \citenamefont {Hirsch}}]{Bastarrachea2017}%
  \BibitemOpen
  \bibfield  {author} {\bibinfo {author} {\bibfnamefont {M.}~\bibnamefont
  {Bastarrachea-Magnani}}, \bibinfo {author} {\bibfnamefont {A.}~\bibnamefont
  {Rela{\~n}o}}, \bibinfo {author} {\bibfnamefont {S.}~\bibnamefont
  {Lerma-Hern\'{a}ndez}}, \bibinfo {author} {\bibfnamefont {B.}~\bibnamefont
  {L\'{o}pez-del Carpio}}, \bibinfo {author} {\bibfnamefont {J.}~\bibnamefont
  {Ch\'{a}vez-Carlos}},\ and\ \bibinfo {author} {\bibfnamefont {J.~G.}\
  \bibnamefont {Hirsch}},\ }\href {https://doi.org/10.1088/1751-8121/aa6162}
  {\bibfield  {journal} {\bibinfo  {journal} {Journal of Physics A:
  Mathematical and Theoretical}\ }\textbf {\bibinfo {volume} {50}},\ \bibinfo
  {pages} {144002} (\bibinfo {year} {2017})}\BibitemShut {NoStop}%
\end{thebibliography}%

\end{document}